\documentclass[10pt,journal,compsoc]{IEEEtran}
\usepackage{ifpdf}
\ifpdf%                                % if we use pdflatex
  \pdfoutput=1\relax                   % create PDFs from pdfLaTeX
  \usepackage{graphicx}                % allow us to embed graphics files
  \DeclareGraphicsExtensions{.pdf,.png,.jpg,.jpeg} % for pdflatex we expect .pdf, .png, or .jpg files
\else%                                 % else we use pure latex
  \usepackage{graphicx}                % allow us to embed graphics files
  \DeclareGraphicsExtensions{.eps}     % for pure latex we expect eps files
\fi%

\graphicspath{{figures/}{pictures/}{images/}{./}} % where to search for the images

\usepackage{url}

\usepackage{microtype}                 % use micro-typography (slightly more compact, better to read)
\PassOptionsToPackage{warn}{textcomp}  % to address font issues with \textrightarrow
\usepackage{textcomp}                  % use better special symbols
\usepackage{mathptmx}                  % use matching math font
\usepackage{times}                     % we use Times as the main font
\usepackage{cite}                      % needed to automatically sort the references
\usepackage{tabu}                      % only used for the table example
\usepackage{booktabs}                  % only used for the table example
\usepackage[utf8]{inputenc}
\usepackage{multirow}
\usepackage{makecell}
\usepackage{float}
\usepackage[usenames,dvipsnames,table,xcdraw]{xcolor}
\newcommand{\fref}[1]{\figurename~\ref{#1}}
\newcommand{\sref}[1]{Section~\ref{#1}}

\usepackage{todonotes}
\begin{document}
%\title{Inviwo --- A Modular Visualization Framework Supporting Visual Debugging}
%\title{Inviwo --- An Interactive Visualization Framework with Visual Introspection}
%\title{Inviwo --- A Visualization Framework\\ with Visual Introspection}
\title{Inviwo --- A Visualization System with Usage Abstraction Levels}
%\title{User and system abstraction layers for visualization systems as realized in Inviwo --- the Interactive Visualization Workshop}

%% indicate IEEE Member or Student Member in form indicated below
\author{Daniel~J\"{o}nsson,
  Peter~Steneteg,
  Erik~Sund\'{e}n,
  Rickard~Englund,
  Sathish~Kottravel,\\
  Martin~Falk,~\IEEEmembership{Member,~IEEE,}
  Anders~Ynnerman,
  Ingrid~Hotz,
  and~Timo~Ropinski~\IEEEmembership{Member,~IEEE,}%
  \IEEEcompsocitemizethanks{
%% insert punctuation at end of each item
\IEEEcompsocthanksitem
D.~J\"{o}nsson, P.~Steneteg, E.~Sund\'{e}n, R.~Englund, S.~Kottravel, M.~Falk, A.~Ynnerman, and I.~Hotz are with Link\"{o}ing University, Sweden.
 E-mail: \{daniel.jonsson, peter.steneteg, erik.sunden, rickard.englund, sathish.kottravel, martin.falk, anders.ynnerman, ingrid.hotz\}@liu.se. 
 The first two authors contributed equally to this work.
\IEEEcompsocthanksitem 
T.~Ropinski is with Ulm University, Germany, and Link\"{o}ping University, Sweden. E-mail: timo.ropinski@uni-ulm.de.
}% <-this % stops an unwanted space
\thanks{Manuscript received November 30, 2018; revised Mars 22, 2019.}}

%\authorfooter{
%% insert punctuation at end of each item
%\item
%D.~J\"{o}nsson, P.~Steneteg, E.~Sund\'{e}n, R.~Englund, S.~Kottravel, M.~Falk, A.~Ynnerman, and I.~Hotz are with Link\"{o}ing University, Sweden.
% E-mail: \{daniel.jonsson, peter.steneteg, erik.sunden, rickard.englund, sathish.kottravel, martin.falk, jochen.jankowai, anders.ynnerman, ingrid.hotz\}@liu.se. 
% The first two authors contributed equally to this work.
%\item 
%T.~Ropinski is with Ulm University, Germany, and Link\"{o}ping University, Sweden. E-mail: timo.ropinski@uni-ulm.de.
%}

%other entries to be set up for journal
%\shortauthortitle{J\"{o}nsson \MakeLowercase{\textit{et al.}}: Inviwo}
\markboth{IEEE TRANSACTIONS ON VISUALIZATION AND COMPUTER GRAPHICS, VOL X, NO. Y, MAY 2019}%
{J\"{o}nsson \MakeLowercase{\textit{et al.}}: Inviwo}

%% Abstract section.
\IEEEtitleabstractindextext{%
\begin{abstract}
%\abstract{
The complexity of today's visualization applications demands specific visualization systems tailored for the development of these applications. Frequently, such systems utilize levels of abstraction to improve the application development process, for instance by providing a data flow network editor. Unfortunately, these abstractions result in several issues, which need to be circumvented through an abstraction-centered system design. Often, a high level of abstraction hides low level details, which makes it difficult to directly access the underlying computing platform, which would be important to achieve an optimal performance. 
Therefore, we propose a layer structure developed for modern and sustainable visualization systems allowing developers to interact with all contained abstraction levels. We refer to this interaction capabilities as usage abstraction levels, since we target application developers with various levels of experience. We formulate the requirements for such a system, derive the desired architecture, and present how the concepts have been exemplary realized within the Inviwo visualization system. Furthermore, we address several specific challenges that arise during the realization of such a layered architecture, such as communication between different computing platforms, performance centered encapsulation, as well as layer-independent development by supporting cross layer documentation and debugging capabilities.
%}
\end{abstract}

%% Keywords that describe your work. Will show as 'Index Terms' in journal
%% please capitalize first letter and insert punctuation after last keyword
}
%\keywords{Visualization systems, data visualization, visual analytics, data analysis, computer graphics, image processing.}

%% ACM Computing Classification System (CCS). 
%% See <http://www.acm.org/class/1998/> for details.
%% The ``\CCScat'' command takes four arguments.

%\CCScatlist{ % not used in journal version
% \CCScat{K.6.1}{Management of Computing and Information Systems}%
%{Project and People Management}{Life Cycle};
% \CCScat{K.7.m}{The Computing Profession}{Miscellaneous}{Ethics}
%}

%% Uncomment below to disable the manuscript note
%\renewcommand{\manuscriptnotetxt}{}

%% Copyright space is enabled by default as required by guidelines.
%% It is disabled by the 'review' option or via the following command:
% \nocopyrightspace

%\vgtcinsertpkg


%\firstsection{Introduction}

\maketitle
\IEEEraisesectionheading{
\section{Introduction}\label{sec:introduction}} %for journal use above \firstsection{..} instead
% MANY VISUALIZATION SYSTEMS EXPLOIT LAYERD STRUCTURE FOR ABSTRACTION
The field of visualization is maturing, and a shift can be observed from algorithm-centric research to application-centric research. While in previous years research has focused on novel visualization methods and algorithms for a particular data type, e.g., volume rendering~\cite{JSYR14}, line integral convolution~\cite{cabral1993imaging}, or tensor glyphs~\cite{kindlmann2004superquadric}, today visualization research also puts emphasis on the solution of specific application-oriented visualization problems in a wide range of domains, e.g., in medical visualization~\cite{Smit17}, engineering sciences~\cite{chen2016visualization}, biological visualization~\cite{le2015cellview}, or astronomy~\cite{BPMRYR15}. Visualization researchers are confronted with new challenges, as in most applications the interplay between different visualization algorithms and the integration of multiple data sources must be considered. Furthermore,  data is typically large and heterogeneous. All these aspects make it challenging to develop interactive visualization applications that are applicable to real-world problems. Over the years, several efforts have been made to provide systems designed to ease the development of  visualization applications. VTK~\cite{schroeder2006visualization} was one of the foundational frameworks for this purpose, enabling visualization application developers to load the data to be visualized and combine several building blocks into a visualization pipeline. Following the success of VTK, many visualization systems have been released to enable application-centric visualization research, e.g., VisTrails~\cite{bavoil2005vistrails}, VolumeShop~\cite{VOLUMESHOP}, MeVisLab~\cite{mevislab}, VisIt~\cite{HPV:VisIt}, VAPOR~\cite{VAPOR}, Voreen~\cite{MRMH09} and Amira~\cite{amira}. While the capabilities of these systems vary, most of them make use of the popular separation of concerns design principle~\cite{reade1989elements} by employing different layers of abstraction. Since the visualization pipeline can readily be modeled through data flow networks~\cite{M13}, many modern visualization systems expose the highest layer of abstraction as a data flow network editor to the application developer, e.g., VisTrails~\cite{bavoil2005vistrails}, MeVisLab~\cite{mevislab} or Voreen~\cite{MRMH09}. This layered architecture approach has been proven successful, but it also comes with a few downsides, which could potentially hamper the visualization application development process.

% ELABORATE ON USAGE ABSTRACTION LEVELS
It can first be observed that visualization application developers have varying needs. Exposing the highest layer of abstraction is often not enough to facilitate the development of complex visualization applications, as in many scenarios it becomes necessary to modify functionality rooted at different levels of abstraction. For instance, besides modifying the data flow network a developer might also want to change the C++ or computing platform specific code for a particular building block. As data flow network editors support rapid development, i.e., the influence of made changes can be directly seen in the visualization, such a development paradigm is also desired at the lower levels of abstraction. Furthermore, as a consequence of such changes, it also becomes necessary to debug and document at these levels. We refer to the possibility to change, debug, test and document functionality at different abstraction layers to \emph{cross-layer development}. We regard our realization of this concept as one of the two main contributions of this system paper.

\renewcommand\theadalign{cl}
\newcommand{\mccc}[1]{\makecell[cc]{#1}}
\newcommand{\tb}{\textbullet}
\begin{table*}[ht]
\centering
\caption{Comparison of major visualizations systems with respect to seamless combination of algorithms implemented on different computing platforms and the support for cross-layer development.}
\label{tbl:comparison}
\begin{tabular}{lc|c|c|c|c|c|c|}
\cline{3-8}
\multicolumn{2}{l|}{} & Inviwo & MeVisLab~\cite{mevislab} & Amira~\cite{amira} & ParaView~\cite{Ayachit2015ParaView} & VisIt~\cite{HPV:VisIt} & VisTrails~\cite{bavoil2005vistrails} \\ \hline
\rowcolor[HTML]{f0f0f0} 
\multicolumn{2}{|l|}{License} & BSD & Commercial & Commercial & BSD & BSD & GNU GPL v2 \\ \hline
% \multicolumn{1}{|l|}{}  & CPU & \tb & \tb & \tb & \tb & \tb & \tb \\ \cline{2-8} 
% \multicolumn{1}{|l|}{}  & OpenGL & \tb & \tb & \tb & \tb & \tb & \tb \\ \cline{2-8} 
% \multicolumn{1}{|l|}{}  & OpenCL & \tb & \tb & \tb  & \tb & \tb & \tb \\ \cline{2-8} 
% \multicolumn{1}{|l|}{\multirow{-4}{*}{Computing Platform}} & CUDA & - & - & \tb & \tb & \tb & -  \\ \hline
\multicolumn{2}{|l|}{Integrated computing platform interoperability} & \tb & --- & --- & --- & --- & --- \\ \hline
\rowcolor[HTML]{f0f0f0}
\multicolumn{1}{|l|}{}  & C++ & \tb & \tb & --- & --- & --- & NA \\ \cline{2-8} \rowcolor[HTML]{f0f0f0}
\multicolumn{1}{|l|}{\multirow{-2}{*}{Interactive development}} & Shaders & \tb & \tb & \tb & --- & --- & --- \\ \hline
%\rowcolor[HTML]{f0f0f0}
\multicolumn{2}{|l|}{Visual debugging} & \tb & \tb & --- & --- & --- & --- \\ \hline
\rowcolor[HTML]{f0f0f0}
\multicolumn{2}{|l|}{Integrated documentation} & \tb & --- & --- & --- & --- & \tb \\ \hline
\multicolumn{2}{|l|}{Visualization pipeline testing} & \tb & \tb & \tb & \tb & \tb & \tb \\ \hline
\end{tabular}
\end{table*}
% ELABORATE ON COMPUTING PLATFORM interoperable
A fundamental problem in the layer abstraction used by modern visualization systems is related to the underlying hardware used. While it is undisputed that modern visualization applications should be interactive~\cite{van2005value}, the size and complexity of data often makes it challenging to maintain interactive frame rates. To deal with this challenge, the ever increasing capabilities of graphics processing units (GPUs) is exploited~\cite{westermann:1998:volRen}, and consequently most current visualization systems, e.g., \cite{bavoil2005vistrails}, \cite{VOLUMESHOP}, \cite{mevislab}, \cite{VAPOR}, \cite{MRMH09} and \cite{amira}, rely on the GPU for both data processing and rendering. This is often done by interfacing with computing platforms, such as OpenCL, CUDA or OpenGL, from within the lowest level of abstraction. While this allows for leveraging the power of modern GPUs, it also comes with the drawback that algorithms cannot be combined with new computing platforms, e.g., an algorithm developed for OpenCL cannot directly be used with an algorithm developed for OpenGL. This is especially problematic when considering the fact that new computing platforms are released rather frequently (2004:OpenGL 2.0, 2006:DirectX 10, 2007:CUDA, 2008:OpenGL 3.0, 2009:OpenCL/DirectX 11, 2015:DirectX 12, 2016:Vulkan). In this paper, we propose concepts enabling algorithms implemented for different computing platforms to be used together seamlessly. 
For example, the image resulting from a rendering using OpenGL can be piped into an image processing algorithm developed using OpenCL. Neither of the two algorithms need to be aware of each other and both algorithms can be programmed using their specific computing platform, which is not possible in other visualization frameworks.
This computing platform interoperability is the second main contribution of this system paper.

% OUR CONTRIBUTIONS
In this paper we propose design principles, and their technical realizations, which support cross-layer development and computing platform interoperability in modern visualization systems.  
%A fundamental principle is to ensure computing platform independence. 
The proposed design principles have been developed and integrated into the Inviwo visualization system, which has been used to implement a range of visualization applications, e.g., volume data exploration~\cite{JFY16}, molecular visualization~\cite{Konig2018}, and material science~\cite{kottravel2017visual}. The proposed concepts are in themselves a contribution beyond the Inviwo system. As visualization systems with cross-layer development and integrated computing platform interoperability will make it possible to use the same system during the entire visualization application development process and gracefully adapt to the introduction of new computing platforms, we foresee that other system developers will find our principles and realization approaches useful.

\section{Related Work}
\label{sec:relatedwork}

% \usepackage{multirow}
% If you use beamer only pass "xcolor=table" option, i.e., \documentclass[xcolor=table]{beamer}

Here, we provide a comprehensive comparison between visualization systems and game engines with respect to their computing platform interoperability capabilities and usage abstraction layers for visualization application development.

% \noindent\textbf{Design patterns} for abstraction exist  for a wide range of problems in the context of software development.
% A more in-depth description of the ones presented here can be found in the book by Freeman et al.~\cite{freeman2004head}.
% The facade pattern provides a common interface to a set of subsystems. 
% This pattern is popular among game engines such as Unity~\cite{craighead2008Unity} to allow support for different computing platforms.
% The proposed system instead uses data abstraction together with the decorator and factory patterns to dynamically add support for new computing platforms.
% Common design patterns in visualization systems are the model-view-controller, which separates data from logic and display, and brushing and linking~\cite{Brushing87}, which allows selection of items in one visualization to be reflected in another one.
% Many visualization systems for example apply the model-view-controller pattern for the visualization pipeline itself~\cite{Ayachit2015ParaView, mevislab,amira,bavoil2005vistrails,MRMH09}.

\noindent\textbf{Visualization systems.} 
The Visualization Toolkit (VTK) by Schroeder et al.~\cite{schroeder2006visualization} is a C++ framework for creating visualizations and therefore requires several usage abstraction levels to be built on top of the framework for it to be accessible to visualization practitioners.
More recently, VTK-m~\cite{moreland2016vtk} has been developed to better support parallel architectures.
Unlike the approach taken in this work, VTK-m provide an abstraction preventing access to the underlying computing platforms. 
Since VTK and VTK-m are lower abstraction level frameworks, they can be integrated into Inviwo and thereby benefit from its usage abstraction levels.

Examples of systems providing usage abstraction levels are ParaView~\cite{Ayachit2015ParaView}, VisTrails~\cite{bavoil2005vistrails}, VisIt~\cite{HPV:VisIt} and tomviz~\cite{tomviz}.
Since VTK does not have integrated computing platform interoperability, the systems building on this API do not have this property either.
Similar to Inviwo, ParaView and VisTrails are general platforms that do not target a specific domain or type of data. 
Out of these, ParaView uses a tree-view to represent the visualization pipeline while Inviwo and many others, e.g., MevisLab~\cite{mevislab}, Amira~\cite{amira},~VisTrails\cite{bavoil2005vistrails}, and Voreen~\cite{MRMH09}, uses an acyclic graph representation.
VisIt primarily targets parallel systems and streaming of data defined on 2D and 3D meshes, while tomviz focuses on transmission electron microscope data. 

There are also many general systems that are not based on VTK.
Amira and MeVisLab are commercial systems targeting life sciences.
Another commercial system is AVS Express~\cite{AVSExpress}, but it does not target a specific domain. 
Voreen was designed for volume rendering and has OpenGL built into its core, which prevents it from integrating computing platform interoperability. 
Forbes et al.~\cite{forbes2010behaviorism} introduced a Java-based framework designed for dynamic data by explicitly separating processing into three interconnected graphs. 
Their graphs involve scene graph logics, data processing and timing operations for interaction and animation.
The work by Forbes et al.~\cite{forbes2010behaviorism} can be seen as orthogonal to the concepts presented here since they could be used together as an additional usage abstraction level with respect to scene graph logics and timing operations.
A range of systems are more specialized, such as FAnToM~\cite{FAnToM}, focusing on topological analysis, VAPOR, focusing on data from earth and space sciences and VolumeShop~\cite{VOLUMESHOP}, focusing on illustrative visualization of volume data.
While most of these systems support multiple computing platforms, none of them allow algorithms implemented for different computing platforms to be seamlessly combined.

\textbf{Visualization systems comparison.} In the following, we will compare several major visualization systems with respect to their computing platform interoperability support and cross-layer development capabilities. The ones selected for comparison, seen in Table \ref{tbl:comparison}, are actively developed and support the whole visualization design process. 
Four concepts were considered when it comes to cross-layer development, being able to develop multiple layers of abstraction at runtime, visual debugging of the data flow network, co-locating code and documentation for higher abstraction layers and testing at functional and visualization pipeline levels, i.e., unit and visualization integration and regression testing. 

As seen in Table \ref{tbl:comparison}, many systems have support for multiple computing platforms but only Inviwo has a concept for seamlessly combining algorithms written for different computing platforms. 
Multiple systems have support for interactive shader editing (Amira, Inviwo, MeVisLab), i.e., changing them at runtime, while only a few provide means for interactively developing algorithms (Inviwo, MeVisLab).

In the software development process, debugging involves locating, identifying, and correcting a defect~\cite{gelperin1988growth}.
Debugging therefore involves everything from printing variables to the console to stepping through code to see where logical errors are made. 
In the field of information visualization, Hoffswell et al.~\cite{hoffswell2015visual} presented techniques for debugging reactive data visualizations. 
Hoffswell et al. focus on showing the state and interaction logic over time as the user interacts with a visualization.
Their techniques can be seen as orthogonal to the ones presented in this work, where the focus is on the debugging process of algorithms and not the the interaction logic.
Debugging of visualization pipelines is supported by Inviwo and MeVisLab.
However, the approach in MeVisLab is slightly different. While Inviwo uses visualization pipelines to provide information co-located with the port location, MeVisLab displays the information in a separate pane.

When it comes to documentation of higher abstraction level components, i.e., processors and modules, it is most common to provide it in separate files. 
For example, MeVisLab requires a .mhelp file for describing fields in their modules.
Here, Inviwo and VisTrails stand out.
VisTrails allows developers to specify documentation in code for their modules (their equivalent to processor) and ports separately, while Inviwo co-locates this documentation for multiple usage abstraction levels.
All of the systems in the comparison provide testing capabilities for both individual functions as well as on visualization pipeline level. 
The main difference is on how easy it is to add the tests. 
Here, we believe that the minimal requirement in Inviwo for adding regression tests, e.g., a workspace with its resulting image(s), can lower the threshold for providing such tests.
A final note is that all visualization systems included in the comparison supports use of scripting languages, such as Python or Tcl.

\textbf{Game engines}, such as Unity\cite{Unity} and Unreal Engine\cite{UnrealEngine}, have lately become viable options for building visualization systems. 
They provide support for multiple computing platforms but do not allow developers to access them.
Sicat et al.~\cite{sicat2018dxr} presented a toolkit based on Unity for creating immersive visualizations.
However, compared to Inviwo their toolkit do not provide usage abstraction levels for editing visualization pipelines.  
In general, the visual debugging support of game engines is more focused on the needs of game developers and not on visualization pipelines. 
Furthermore, the game engines' focus on game developers can limit their use for practical visualization problems. 
For example, Unity pre-processes meshes to optimize them for rendering, which might not be desirable for scientific visualization purposes.

%% -----------------------------------------------------------------------------------------------------------
\section{System Requirements}\label{sec:systemrequirements}
\begin{figure}
	\centering
	\includegraphics[width=1.0\linewidth, trim=0 0 170 0, clip]{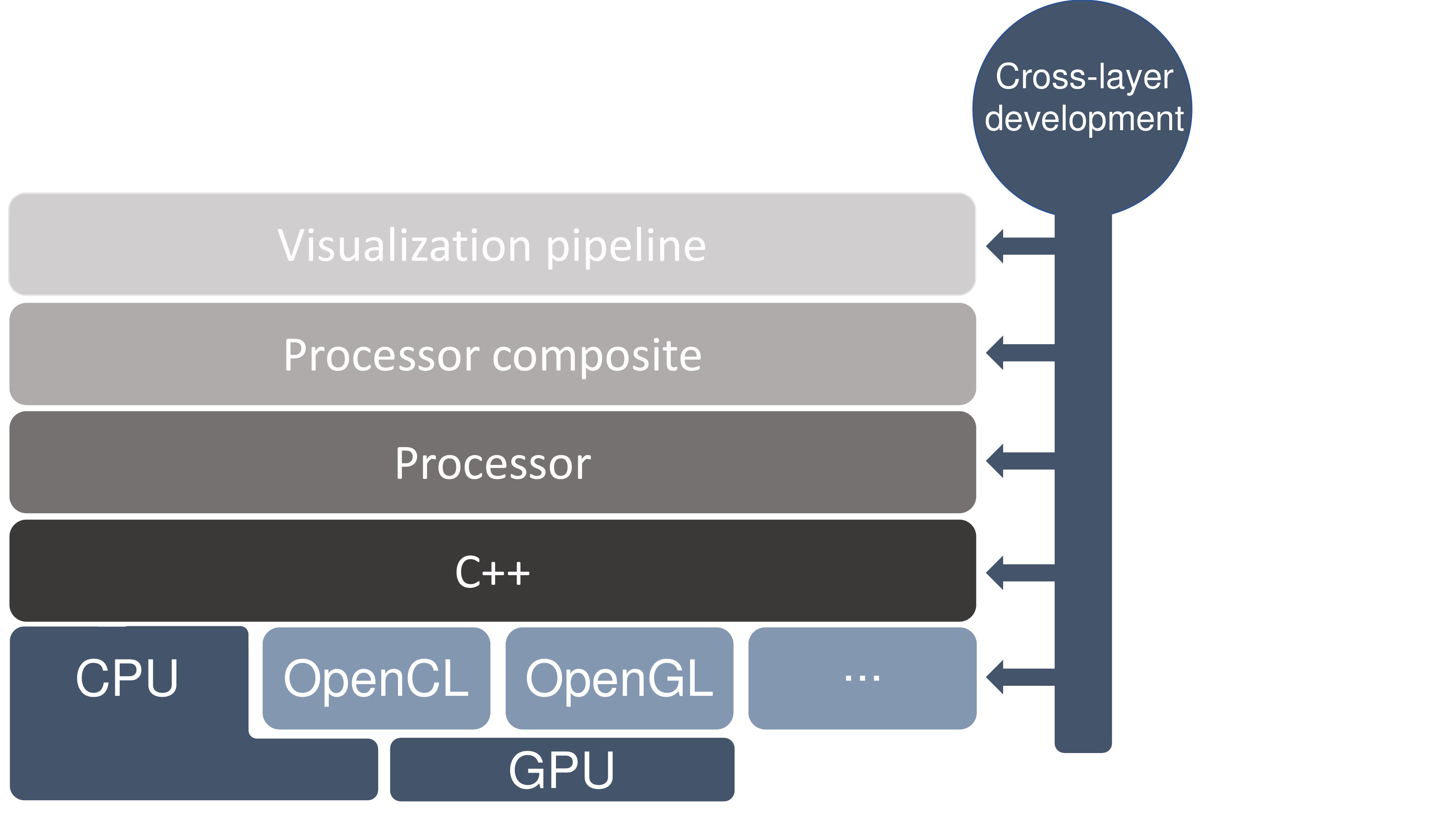}
	\caption{\label{fig:AbstractionLayers} Illustration of the layers of abstraction in our visualization system.
    The different layers of abstraction allow access to functionality for different usage levels and parts of the visualization application design process. 
    As an example, our Python integration allows interactive development across layers through editing of the visualization pipeline or definition of new processors. 
    %All levels can be accessed though python scripting and can be seen as an entry point to writing lower level code.  
    }
\end{figure}

Visualization systems commonly use multiple layers of abstraction to realize an abstraction-centered design.
%The final tailored application serves as a top abstraction layer.
%The basis for the tailored application is a high level visualization pipeline design layer.
As illustrated in \fref{fig:AbstractionLayers}, the top abstraction layer is the visualization pipeline, which act as a layer of abstraction for processor composites, i.e., a group of functional units processing data flowing in the visualization pipeline.
Each processor is in turn acting as an abstraction layer for C++ code. 
The lowest abstraction layer provides computing platform level access.
Having multiple layers of abstraction poses several challenges that we address in this paper.
Foremost, how can a cross-layer development process be provided, where the developer can use concepts from low abstraction layers, i.e., debugging and testing, on the higher abstraction layers. 
Also, which parts of the lower abstraction layers should be exposed on higher levels.
Critical aspects that we consider here are performance and development speed, it must be ensured that the access across abstraction layers does not hamper performance and that it helps in creating tailored visualization applications.
Another challenge is communication within the computing platform layer, for example how to deal with the number computing platform combinations possible.

Application oriented research requires a new level of engineering and our aim is to \textit{supply a sustainable system that aids researchers in all stages of the visualization application design process.} %\textit{reduce the amount of work necessary to realize this type of research and support the developer in debugging the increasingly complex applications}.
We first describe the requirements for the usage abstraction levels and then the computing platform interoperability.

A requirement for visualization systems with usage abstraction levels is that they support rapid development at all levels ranging from application tailoring to computing platform coding. 
From a developer's perspective, this process involves coding, debugging, documenting and testing.
Thus, the system should support interactive C++ coding, shader editing and debugging of data flow networks.
Documentation of algorithms is read by other developers and visualization composers, meaning that both groups must have access to it on the level they are using the system.
Testing must be supported on both algorithm and visualization pipeline levels.
From a visualization composer's perspective, the system must support visual creation and editing of visualization pipelines as well as the possibility to create a tailored visualization application. 

Interaction is key in visualization systems.
A computing platform interoperable visualization system must therefore provide access to native computing platform capabilities to allow processing of large and complex data at interactive frame rates. 
In addition, research is commonly at the forefront of technology usage, which means that a visualization system used for research cannot abstract away access to computing platforms.
Thus, such a system must instead provide access to the lowest technical layer.
For algorithms to remain portable across computing platforms, the system must provide ways of seamlessly converting data between computing platforms, i.e., transfer data between memory locations.
While it is possible to manually convert data from one computing platform to another one, it requires knowledge about both computing platforms, and thus does not scale when later introducing new computing platforms. 
%A computing platform interoperable system allowing access to the underlying computing platforms must therefore be able to do this conversion automatically. 
Furthermore, since this is a traveling salesman problem where the highest performance path from one computing platform to another one must be found, it is not feasible to solve this combinatorial problem manually even for a small number of computing platforms.
To make things more complicated, different computing platforms also have support for different types of specialized data, such as buffers or textures. 
Mapping between different types of data occurring in different computing platforms is therefore an additional challenge that needs to be considered.
%This cannot be expected to be done manually by a developer sothe number of combinations between computing platforms grows by $\sum_{1}^{n}(n-1)$, where $n$ is the number of different platforms.
%In practice, there are also different types of data, e.g., buffer, 2D texture or 3D texture, and, in addition, it is not always possible to convert directly between two compute platforms.
%This means that that the number of possible combinations quickly become large even for a small number of compute platforms. 
%Thus, it is unfeasible to expect a developer to implement all different possible combinations.
%The system must be able to solve this combinatorial problem given at least one conversion path from a compute platform data type to another one.

Concepts addressing the requirements above are presented in the following two sections.

\section{Usage Abstraction Levels}

% This work uses a layered architecture structure to meet the requirements described in the previous section.
% This layered architecture structure allows both multiple levels of abstraction and cross-layer access. % to meet the demands of different usage levels.
% An overview of the structure is first provided before discussing key components allowing cross-layer development.
Supporting interactive coding, shader editing and debugging requires a system allowing development to be performed across layers of abstraction.
This section will define key concepts for such a system meeting the previously presented requirements.
First, we describe ways of achieving interactive development. 
This is followed by three concepts enabling cross-layer development in terms of debugging, documenting and testing. %, are discussed in more detail. %, a serialization-based concept and a visual debugging concept.. 
%Two key concepts enabling cross-layer development will be then be described here, a serialization-based concept and a visual debugging concept.. 
%The presented concepts and the full range of usage abstraction levels have been implemented in Inviwo, as described below.

\subsection{Interactive Development}
High level development means using a visual representation for creating and changing visualization pipelines.
In order to allow simultaneous high level editing and low-level C++ coding, the system must recompile and reload the changed parts while the visual editor is running. 
This concept must be realized through a modular system design, i.e., a plug-in system where dynamic libraries can be unloaded and loaded again to reflect the changes made.
Medium usage abstraction level editing, i.e., modifying shaders used by computing platforms, or scripting, can be performed at runtime without unloading or loading dynamic libraries.
Thus, interactive development at this level can be realized through either providing a user interface for recompiling the source code or observing the sources of origin, e.g., text files, and automatically recompile the source code when it changes.
While the presented interactive development concept is seemingly simple, it allows the user to access  multiple layers of abstraction at the same time and thus address the rapid development requirements of the system.

\subsection{Visual Data Flow Debugging}
Debugging parallel operations is difficult since there are many outputs occurring at the same time.
Readily available debugging tools can help developers by allowing the execution of a given thread to be inspected.
Inspecting one thread at a time can be tedious and is most effective when the error has been narrowed down to a specific part in the code. 
Visually inspecting output of parallel operations can  help the developer to quickly obtain an overview and thereby identify the source of an error.
This can be seen as debugging on a higher level of abstraction, giving quick access to information at lower abstraction layers.
% Additionally, the algorithms running on CPU and GPU might be tightly intertwined in a single application.
% The Inviwo framework supports the developers with tools for debugging such CPU and GPU-based algorithms by means of visual debugging.

This cross-layer debugging is provided by a concept we refer to as port inspection.
A port inspector allows the contents of a processor's port to be viewed. % by means of applying a data flow network that takes such a port as input.
By applying this concept in the visualization pipeline  editor, it allows for step-by-step debugging of the data-flow network.
For example, hovering or clicking on ports, or their connections, shows debug information about data originating from their associated outport.
We can not only inspect data on a higher abstraction level, we can also create the debugging information on a high level.
For this purpose, we use a visualization pipeline to provide the debugging information for a particular port type.
For example, a volume data port, containing a stack of images, can use a predefined visualization pipeline for slice-based inspection.
Thus, the presented concept enables cross-layer visual debugging of data flow networks.

\subsection{Documentation}
API documentation is more or less standardized nowadays using for example the Doxygen system. 
However, this documentation is intended for developers and is not readily available for visualization pipeline editors. 
Providing documentation at the visualization pipeline editor level can be done in a wide variety of ways.
For example, by presenting documentation written in separate files, such as xml, or html.
While this makes it possible to use tools suited for text and layout, it also makes it harder to maintain and follow from a developer's perspective.
For systems where low maintenance overhead and ease of use is important, we suggest to integrate this documentation procedure into the API documentation.
This allows developers to write documentation once but show it in both API-documentation and to visualization pipeline editor users.

While developers are used to textual documentation, pipeline editors often desire richer information.
Thus, the documentation must be presented appropriately at the different usage abstraction levels. 
In practice, our concept involves extracting the documentation of processors from the source code and augmenting it with the visual representation used in the higher abstraction layer.   
The augmented visual representation of the processor contextualizes the documentation and can be generated automatically since they are already used in the higher abstraction layer.

The documentation concept described above not only makes it easy to maintain, due to its co-location with the actual code, but also provides information on how to use a processor at multiple usage abstraction levels and thereby address the system requirements. 

\subsection{Testing}\label{sec:testing}
Unit testing is commonly used for testing low abstraction level code, but can also be used for testing an entire visualization pipeline. 
However, unit testing requires programming skills and therefore cannot be created in the higher usage level abstraction.

Our testing concept allows tests of visualization pipelines to be created on a high usage abstraction level, i.e., through the visualization pipeline GUI editor.
Once the visualization pipeline to be tested has been defined its results before and after a change can be compared.
However, comparing results of a visualization pipeline presents challenges that our concept takes into account.
Here, one approach is to generate reference results on demand, i.e., never store them, and compare the new results with the generated references each time a test is run. 
We have found this approach to be problematic due to two reasons.
First, the results can vary between different hardware, but this, possibly incorrect behavior, will never be detected since the same hardware is used for generating reference and comparison images. 
Second, it can be resource demanding since it requires having, or building, multiple versions for generating the reference data. 
Thus, reference results should instead be generated once  when the test is defined.
To deal with the differences in results across hardwares we allow specification of an error threshold such that, in cases an investigation has found that this indeed is the cause of the error, the test can be allowed to pass.

The presented testing concept has addressed the challenge of exposing testing at a high usage abstraction level, and thus enables the system to support testing across layers of abstraction.

% abstraction levels 
% Address both usage abstraction layer and compute platform interoperable
% (architecture)
% Answer why
 %levels, and cross-level visual debugging. 

% We propose a way of serializing across layers.
% The top level allow copy-paste and undo-redo operations.
% The second level allow a processor to be an abstraction of a processor network. 

\section{Computing Platform Interoperability}\label{sec:computePlatformCommunication}
\begin{figure}[t]
  \centering
  \includegraphics[width=1.0\linewidth, trim={40 145px 20 0}]{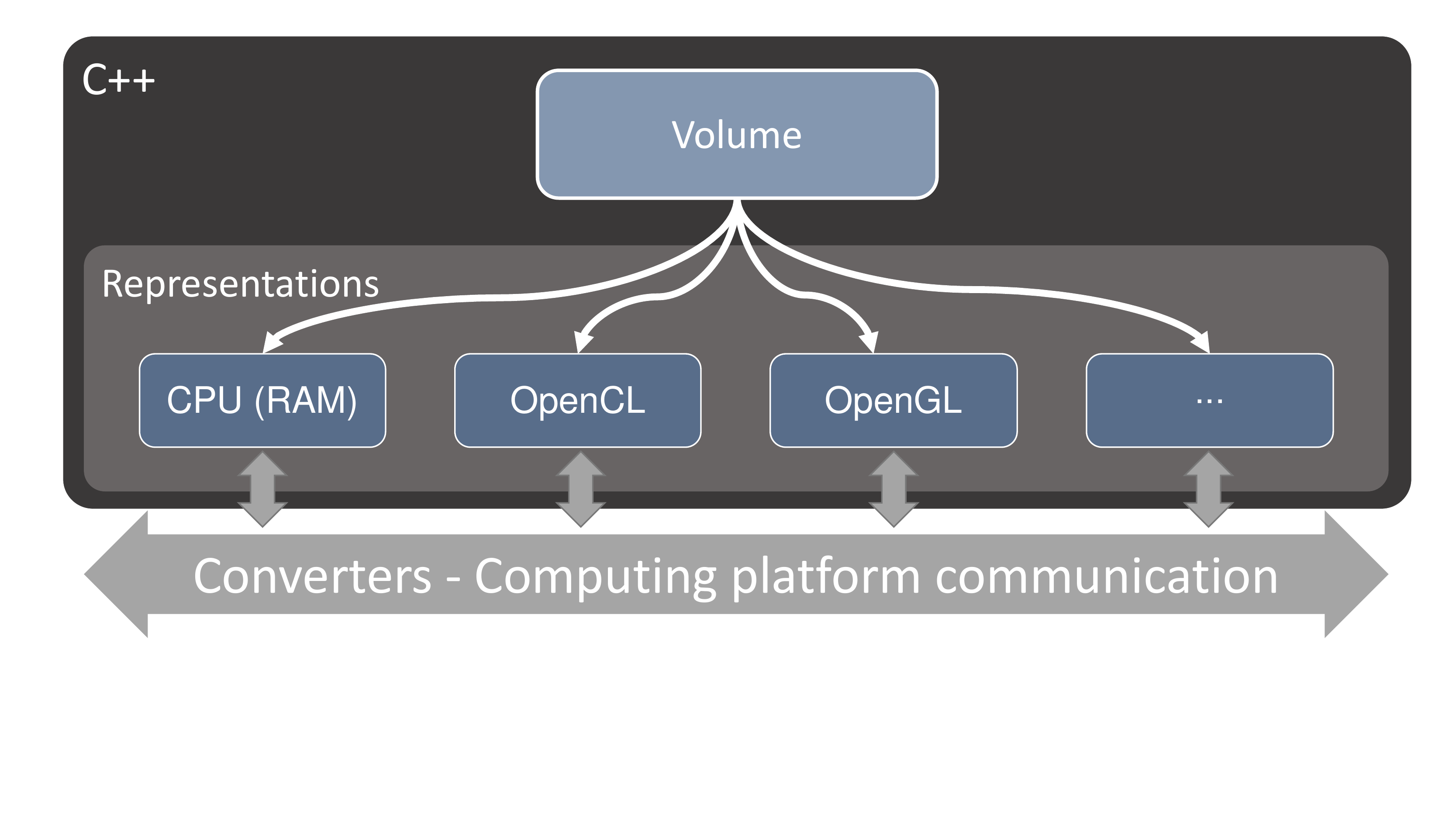}
  \caption{%
    Illustration of the computing platform communication in the C++ usage abstraction level for a volume data object. 
    The Developer requests representations of the object but data transfer is performed on a layer of abstraction underneath data objects, thus never visible to the developer. 
    Representations are mapped to types supported by hardware, i.e., buffers and textures, for performance reasons. 
  }%
  \label{fig:ComputingPlatformCommunication}%
\end{figure}

The lowest abstraction layer in our system is data abstraction.
This layer provides an interface for accessing data on a per computing platform basis without knowing about the other computing platforms, thus enabling developers to utilize platform specific features but still combine algorithms written for different computing platforms.
%The network editor has two modes, developer mode, used for designing visualization pipelines, and application mode, used by practitioners.
% The developer mode allows creation and editing of visualization pipelines, tuning of parameters and selecting which ones to show for practitioners.
% The application mode hides the data flow network and developer parameters and thus act as the top layer of abstraction.

%The following sub sections describe data transfer in the compute platform layer and cross-layer development for serialization, debugging, documentation and testing. %, serialization at different
Designing a computing platform layer of abstraction requires considering both performance and sustainability.
For example, the OpenGL computing platform has specialized data types for textures which must be taken into account to not loose performance.
Therefore, we focus on creating a layer of abstraction for the data, not for all computing platforms. 
In other words, our concept allows developers to fully utilize the computing platforms on the premise that they request data for the computing platform they are utilizing.
While this might seem like a too narrow abstraction, it is fitting to visualization pipelines since they operate on data flows and each processor is in this way isolated from the other ones.
%Therefore, an abstraction layer for accessing data can be used.
This abstraction layer must provide a common interface for data objects flowing in the visualization pipeline.
One way of providing such an abstraction is to disallow direct usage of the underlying computing platform memory location and instead provide a wrapper exposing common functionality, similar to how for example the Java programming language operates.
Clearly, this prevents developers from using computing platform specific features.
Another alternative is to provide an interface for requesting the data for the desired computing platform.
Given that data can be transferred to the requested computing platform, it means that the developer only needs to know about the computing platform for which data is requested for.   
%The upside is that different compute platforms can be used, even for the same data.
The downsides of this approach are a more complicated data interface for developers, due to the abstraction, and that data might need to be transferred when used by different algorithms.

%The second option was chosen since it allows access to compute platform capabilities. 
%Data objects are given a layer of abstraction with respect to accessing the memory location of the underlying data for each compute platform.
%Thus, access to data is only provided when explicitly requested for a given compute platform.
%The system transfers the data, if necessary, to the requested compute platform.
%This means that an algorithm only need to know about the compute platform it is working with and therefore use all of its features.
Our design consists of a general data interface exposing functions for requesting data for arbitrary computing platforms.
Examples of data types utilizing this interface are buffers, images and volumes. 
Succinctly, the computing platform data representation is simply referred to as representation.
The three most important objects of importance in our design are illustrated in \fref{fig:ComputingPlatformCommunication} for  volume data.
In this figure the data interface provides an abstraction for data access, the representation manages the data, and the converter transfers data between different computing platforms.
A representation is created lazily upon request. 
However, once requested it is cached in order to reduce the number of data transfers necessary.
As the representations are typically not on the same device, i.e., RAM and video memory, caching is in general a good trade-off.
A list of used representations by the data object must be stored.
It is necessary to know which representations in the list that are up to date.
For this purpose, a representation must be requested with either read or write access.
Write access means that other representations must be updated from this representation when requested.
The observant reader has identified two types of representation operations, one for creation and one for updating.
It is desirable to separate these two operations since creation is generally slower than updating.
Therefore, the only two tasks of a converter is to create or update a representation from another one.

In practice, computing platforms such as OpenGL and OpenCL have special data object types for which hardware accelerated operations are supported. 
These need to be considered to obtain interactive frame rates.
Based on currently existing hardware we have identified three types of fundamental data objects. 
Buffers, which can store arbitrary data, 2D textures storing image data and 3D textures storing volume data.
This separation ensures that hardware accelerated filtering operations can be utilized on GPU computing platforms.
Moreover, while the computing platforms have evolved, these underlying types have remained the same and due to their ubiquitous usage it is not likely that they will be removed in the near future when new hardware is developed.
Each of these fundamental types requires one converter per combination of computing platform.
In order to form an optimal conversion path when requesting data, all combinations between the existing computing platforms must be considered.
Thus, it quickly becomes unfeasible to manually implement all possible converter combinations.
This challenge is addressed by algorithmically creating converter packages from the individual converters provided by the developers.
One such package contains a path to transfer data from one representation to another. 
These packages can be sorted in order of performance and be selected at runtime.  
In practice, this algorithmic converter path creation means that a developer only has to implement one converter between the new and an existing representation for it to function.
Specifying more data transfer combinations can increase performance, such as in the case of shared data.

More complex data structures, such as meshes containing buffers of vertices and other attributes, can utilize the fundamental types provided as building blocks and thereby gain the same properties. 
The presented concept improves the sustainability of a visualization system while still allowing access to native computing platform capabilities. 
The presented usage abstraction levels and computing platform interoperability concepts have been implemented in Inviwo. Details of these implementations are provided in the following section.
\begin{figure}
	\centering
	\includegraphics[width=\linewidth, trim={140 30px 140px 0px}, clip ]{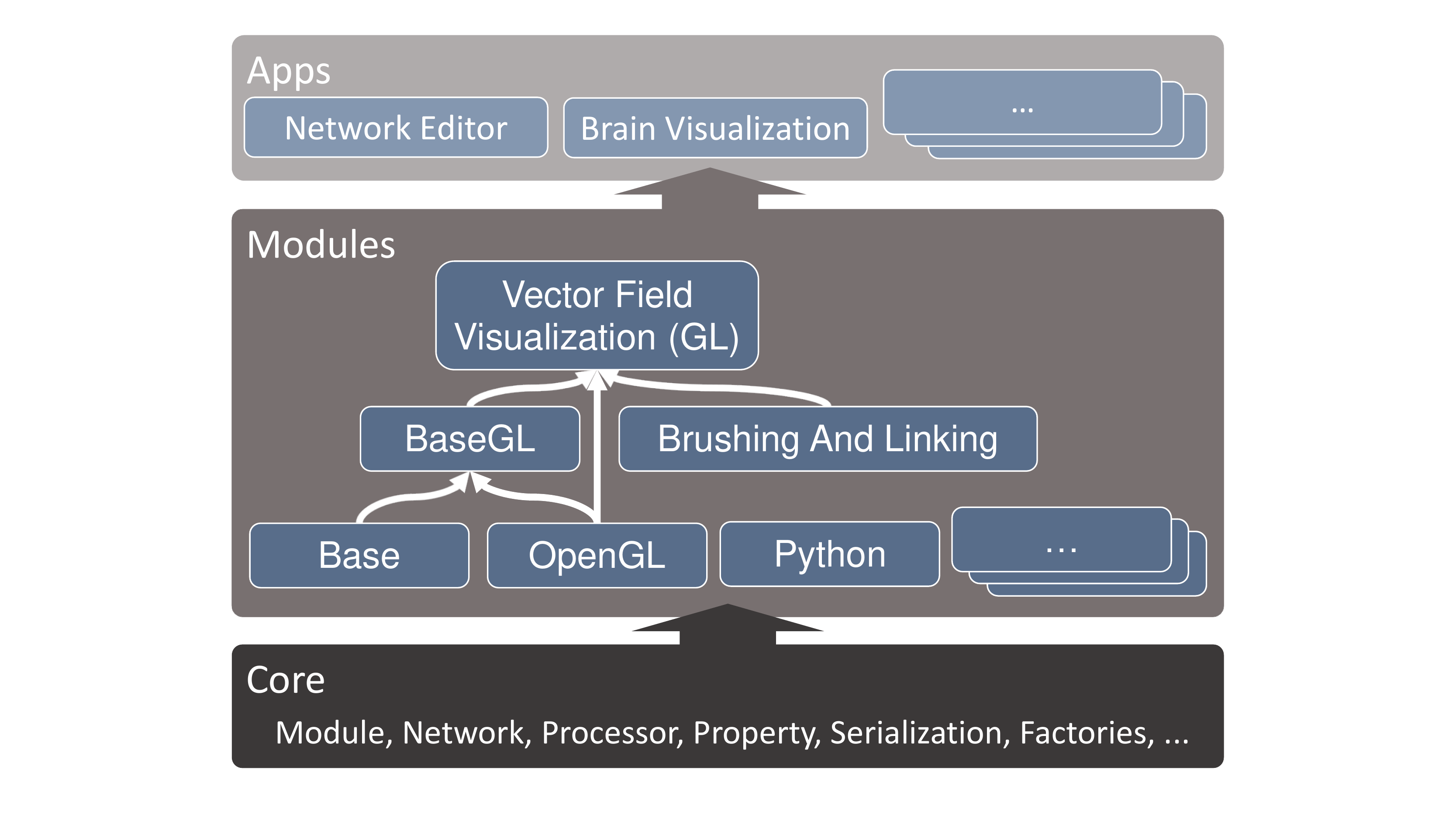}
	\caption{\label{fig:architecture}The Inviwo architecture is divided into three main components. 
    The core provides, for example, processor network functionality, saving/loading mechanisms, and interfaces for extending the framework. 
Modules bring in new functionality using the interfaces and factories provided in core, or by exposing external libraries.
Applications expose the processor network in various ways, for example using graphical interfaces for editing the network or tailored design for brain activity visualization. 
}
\end{figure}

\section{Realization in The Inviwo Framework}\label{sec:inviwosystem}
The previously described concepts have been implemented in the Inviwo system.
We will first provide an overview of Inviwo and describe how its different parts relate to these concepts.
Then, we will discuss how the usage level abstractions and computing platform interoperability are realized in Inviwo.  

% Visualization involves an interactive pipeline of data processing, with which can be interacted by changing algorithm parameters, before interpreting the output on the screen.
% Each application making use of visualization often have unique needs in terms of the set of algorithms to apply on the data, parameter settings or views to be analyzed.
% It is clear that one set of algorithms cannot be applied to all applications.
% The Inviwo system therefore focus on allowing tailored visualization pipelines to be rapidly created.

Inviwo can be divided into three main components illustrated in \fref{fig:architecture}.
First is the core, containing functionality for evaluating a data-flow network, fundamental data structures, such as images, volumes and meshes, as well as interfaces and factories for adding additional functionality.
The core has no dependencies on any GPU platform but provides the foundation for computing platform communication.
Second is the module system, which is used to extend the core with new functionality at runtime in a plug-in based manner. 
%Note here that Inviwo uses the term \textit{module} in a plug-in like manner, not for a functional unit as previously used and defined by Kenneth Moreland~\cite{M13}. 
A new module can, for example, add support for data readers, computing platforms or new visualization algorithms.
Third is the visual editor in which visualization pipelines can be created, parameters can be tuned and scripting can be edited. 
The visual editor provides access across the layers of abstraction and supports interactive C++ coding, scripting and shader editing.
It has two modes, developer and application modes. 
The developer mode allows data flow creation as well as parameter editing, while the application mode hides the data flow network and only shows a subset of parameters determined in the developer mode.
Thus, a tailored visualization application can quickly be created.
% The visual editor has been equipped with powerful features to help practitioners and developers quickly create, edit and visually debug the data-flow network.

The following sub sections will first provide a brief overview of Inviwo, then describe how the usage abstraction levels are implemented followed by the details on how algorithm interoperability is achieved.  % encapsulates the visualization pipeline concept, the metaphors used for algorithms and data and how data is delineated from parameters in an algorithm.

%% -----------------------------------------------------------------------------------------------------------

\subsection*{Inviwo Overview}\label{sec:InviwoCore}
\begin{figure}[t]
	\centering
	\includegraphics[width=\linewidth]{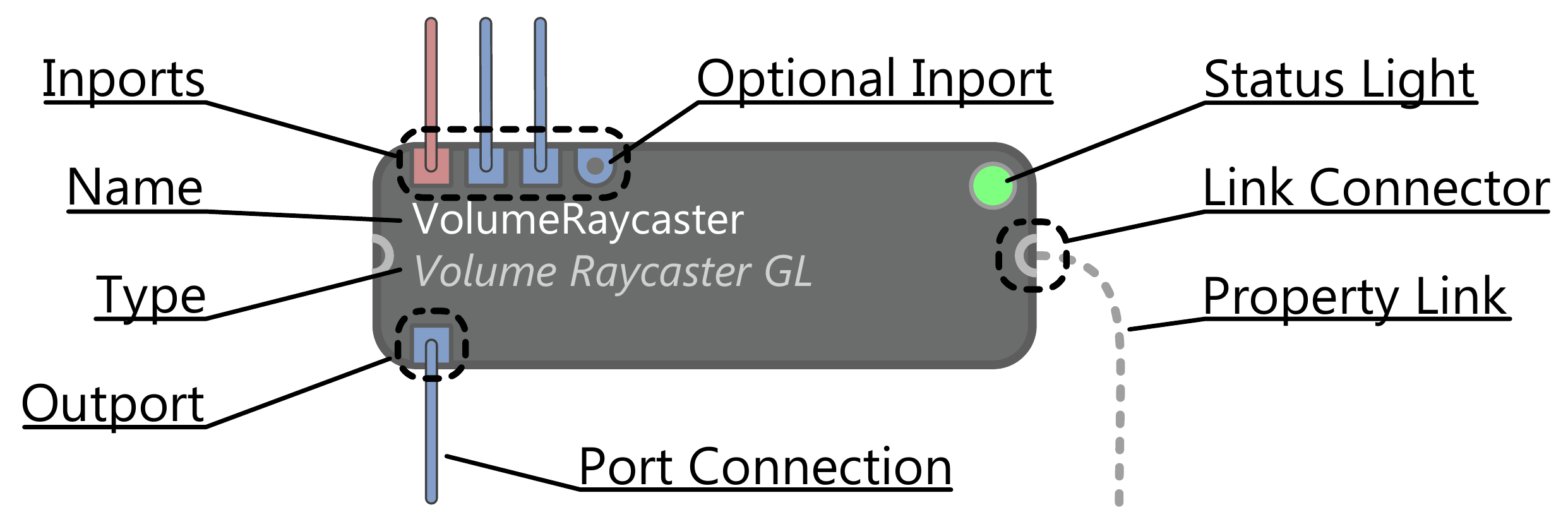}
	\caption{\label{fig:processor} 
	Visual representation of a processor, the encapsulation of an algorithm in Inviwo. 
	Data is flowing from the top into the inports of the processor. 
	The resulting output data continuous its flow through the outport.
	Synchronization of properties between processors are represented through property links.
	}
\end{figure}
Inviwo is built around a central concept of a data flow graph where each node in the graph represents a functional unit.
The edges in the graph represent data flowing into and out of the functional units.
A functional unit, its input and output data and its parameters are encapsulated by a processor.
A processor has inports, encapsulating the input data, outports, encapsulating the output data, and properties, encapsulating parameters.
Guidelines for how to best encapsulate a functional unit into a processor are provided in the appendix of this paper.
A group of processors can be combined in the form of a processor composite and thus provide an abstraction for a more complex task. 
The graph of connected processors and processor composites is referred as the processor network, which is equivalent to a visualization pipeline.

Data can be raw data in memory, geometries, volumes, images, data frames or any other arbitrary data structure. 
Note that while the types provided by Inviwo generally use the data interface presented in \sref{sec:computePlatformCommunication}, there is no such requirement on the port data.
This means that data types in external libraries can easily be integrated and benefit from the usage abstraction levels, even though they will not gain the computing platform interoperability of the data types provided by Inviwo.
Figure~\ref{fig:processor} shows the visual representation used for a processor and its components. %of abstraction how this algorithm metaphor is represented visually.
The processor itself is represented by a box, where input data is flowing from the top into the inports and the output data is exiting in the bottom through the outport.
A processor is evaluated when all its inports, and at least one outport, are connected.
The low abstraction layer concept of function overloading in programming is captured using optional inports, which do not need to be connected for the processor to evaluate.

\subsection{Usage Abstractions For Interactive Development}
The implementations of interactive development in Inviwo can be categorized into three usage abstraction level categories.
The high usage abstraction level is based on graphical interfaces, the medium usage abstraction level contains scripting and shader editing while the lowest usage abstraction level is based on C++ coding.
Next, we describe how the developer can move between these categories and achieve interactive development in Inviwo.

\subsubsection{High Usage Abstraction Level Editing}
\label{sec:networkeditor}
\begin{figure}[t]
	\centering
	\includegraphics[width=\linewidth]{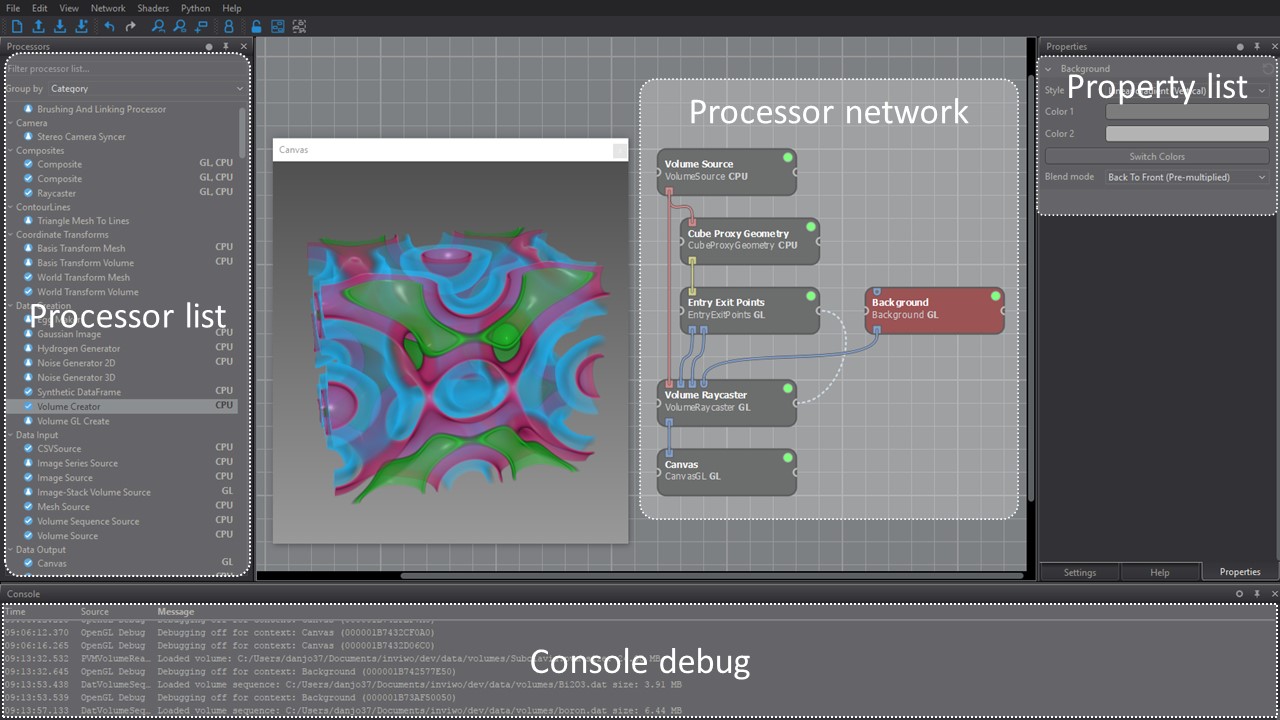}
	\caption{\label{fig:screenshot} Screenshot of the Inviwo network editor for rapid visualization pipeline creation. 
    The processor network in the center loads and visualizes the charge density of the chemical boron element. 
    New processors can be dragged from the processor list and dropped into the processor network, where a visual representation of the processor and its components will be shown. 
    Pipelines are created by connecting processor ports in the network. 
    Properties exposed by the selected processor are automatically shown in the property list to the right. }
\end{figure}
%\subsection{Data flow vs links (Data vs properties).}
% \begin{figure}
% 	\centering
%  	\includegraphics[width=\linewidth]{figures/boron_annotated_minimal}
% 	 \caption{\label{fig:dataflow}The processor network has three orthogonal concepts for information transfer. Data is propagated downwards in the network, from outports to inports, events are propagated upwards from inports to outports, and properties are synchronized between processors using property links.}
% \end{figure}

Visualization pipeline abstraction level editing is performed in the processor network editor (network editor), seen in Figure~\ref{fig:screenshot}. 
The network editor is an application for visually building, editing, debugging and running processor networks.
It is a key component as it provides means for cross-layer development by allowing the developer to seamlessly move between visually editing the visualization pipeline, scripting in Python and coding in C++.

%It is a key element for collaborating with domain experts, where visualization pipelines can be created from existing building blocks and parameters can be tuned together with domain experts.
The network editor uses the drag-and-drop metaphor to include processors in the network. 
Processors added by modules can be selected from a list and dropped into the processor network, whereby a graphical element, exemplified in Figure~\ref{fig:processor}, is created to represent the selected processor in the network. 
Its inports and outports are displayed as colored squares laid out in rows at the top and the bottom of the element, respectively. The color of the ports reflect the type of data they manage. Images are blue, volumes are red and meshes are yellow. 

% An inport and outport is connected by selecting the outport, dragging a connection, and releasing it on the inport. 
% The connection appear as a line during the drag operation, which will be highlighted in red if the outport does not support connection with the inport, i.e., their corresponding data does not match. 
The network is automatically evaluated while building the pipeline and the results can be inspected either in a canvas or through visual debugging (see \sref{sec:IVWVisualDebuggin}). 
Selecting a processor in the network shows graphical user interfaces (GUIs) for editing all of its properties, see the Property List in Figure~\ref{fig:screenshot}.
The same property type can have different visual representations, which we refer to as property semantics. 
For example, four floating point values can be represented by four sliders or as a color. 
The property semantics can be set at different usage abstraction levels, i.e., in code or visually using the property's context menu. 
%Figure CreateSemanticsExampleFigure shows some examples of different semantics supported in the application.
Multiple processors can be aggregated into one processor, which  thereby  enable high usage abstraction level access to the third layer of abstraction, processor composites. 

Properties in the network can be linked (synchronized) between processors in a similar, but orthogonal, way to connections between ports.
A dotted line between link connectors of two processors indicate that one or more of their properties are linked. 

This high usage abstraction level editing enables interactive development of a visualization pipeline.
However, the network editor also integrates with the medium and low usage level abstractions as described next.

\subsubsection{Medium Usage Abstraction Level Editing}
%Interactive development can also be performed through the use of scripting languages. 
Inviwo has integrated support for the widely used Python scripting language. 
The Python integration in Inviwo can be used in several ways. 
Firstly, it is possible to perform batch processing using an integrated Python editor in the processor network editor.
As an example, camera parameters can be scripted for performance benchmarking.
The integrated Python editor thereby enables the developer to seamlessly move between the high and medium usage abstraction levels. 
Secondly, data can be transfered back and forth to Python for computations within a processor.
This is particularly valuable due to the rich data processing capabilities in Python, but also means that algorithms available in Inviwo can be used in Python.
The Python data transfer is typically used within a processor, but the script can be exposed through a property which thus enables high usage abstraction level editing. 
Third, a processor and its processing can be defined entirely in Python. 
This is advantageous for developers more comfortable with Python, and of course inherits the benefits of working with a scripting language while still taking advantage of the usage abstraction level concepts in Inviwo. 
Finally, we point out that Python is still integrated through a module, i.e., it is not included in Inviwo core, which demonstrates the modularity of Inviwo.

Interactive shader editing in the network editor is also supported by observing all shader files.
Saving a shader file will notify the network editor, which will invalidate the processor network, causing it to update the output. 
Similiar to Python scripting, some processors also expose shaders through properties. 
Changing a shader in this case invalidates the property, which also causes the processor network to update.

\begin{figure}[!b]
  \centering
  \includegraphics[width=0.8\linewidth]{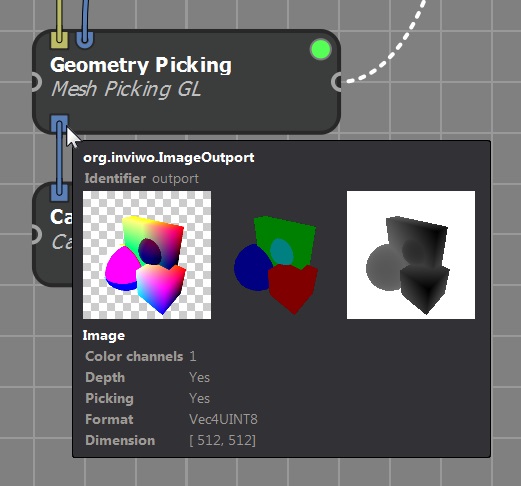}
  \caption{%
    Example of visual debugging of a visualization pipeline.
    In this instance, a port inspector shows the rendered color layer, the picking buffer, and the contents of the depth buffer. 
    Details, such as data format and dimension are, displayed below the images. 
  }%
  \label{fig:portinspector}%
\end{figure}
\begin{figure*}[!ht]
	\centering
	\fbox{\includegraphics[width=0.9\linewidth]{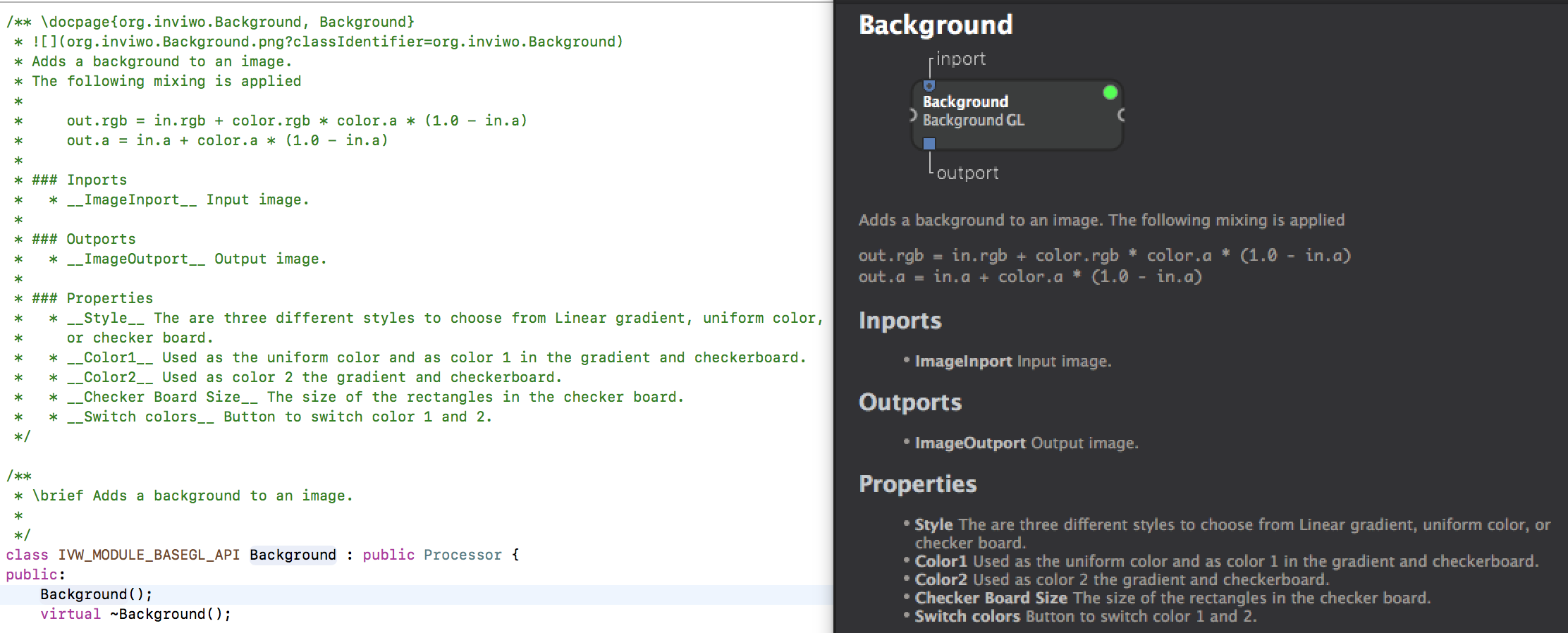}}
 \caption{\label{fig:backgroundDocumentation}
Inviwo processor documentation, based on Markdown and doxygen, serves as both API (left) and end user documentation (right). 
The visual processor representation seen at the top in the right image is automatically generated by the Inviwo documentation system.}
\end{figure*}
\subsubsection{Low Usage Abstraction Level Editing}
The Inviwo system is based on a modular structure, which is the foundation for achieving interactive development. 
All modules depend on the core and, in addition, they can depend on other modules or external libraries, see \fref{fig:architecture}. 
Dependencies on external libraries occur when integrating their functionality into Inviwo. 
The module system allows new functionality to be included without introducing any dependencies into the core system. 
For example, the OpenGL module adds OpenGL computing platform support to the Inviwo system. Other modules can depend on the OpenGL module to build on those capabilities. Inviwo takes care of managing the dependencies between the different modules.
When possible, external library source code should be brought into the build system to ensure version compatibility in deployed applications.

Interactive coding is enabled by allowing modules to be loaded at runtime when they change, also known as hot reloading. 
The network editor observes all module shared library files for change, meaning that the developer can recompile the module while the application is running and immediately see the effects of the change.
The processor network is serialized, modules are reloaded, and the network is deserialized again.
It should be noted that the developer has the responsibility to ensure that a module can be unloaded and loaded repeatedly, for example by not making breaking API changes, as the system cannot regulate this. 
A module version system ensures that modules of correct version are loaded at runtime. 
Modules are only loaded if they are linked to the correct Inviwo core version used by the application. Thus, it is only necessary to bump the module version if the module is released in between Inviwo core releases.
This version dependency means that module developers rarely need to care about the module version.

In essence, the presented interactive development implementation allows developers to rapidly move between a low usage abstraction level, i.e., editing the code, and a high usage abstraction level, i.e., visually editing the visualization pipeline.

\subsection{Visual Debugging}\label{sec:IVWVisualDebuggin}

Inviwo provides port inspectors based on processor networks for three types: meshes, images and volumes.
Modules can add more port inspectors if desired.
Events are forwarded to port inspectors meaning that interaction is possible.
As an example, the volume port inspector, showing a series of slices, allows for changing the current slice using the scroll wheel of a mouse.
Since the pipelines created for port inspection by design are computing platform interoperable, it is possible to use the port inspectors on data residing in another computing platform's memory.

Creating an entire processor network for less commonly used port types can be to high of a threshold. 
Inviwo therefore also allows port debug information to be created using C++ template traits for the port type.
The port template trait output debug text, expected in HTML-format, will be displayed together with the results of the data flow port inspection, if existing.
This enables visual debugging information to be created at both low and high usage abstraction levels.

Figure~\ref{fig:portinspector} depicts the inspection of an image port as realized in the Inviwo system.
It shows the output images generated or forwarded by the respective processor, including additional information such as image format and dimension.
A default implementation for ports displays the name of the port, meaning that the port inspection can be applied to all ports and port connections without any efforts from a developer.

%\subsubsection{Serialization At different Levels}
% The undo-redo serialization concept described in section \sref{sec:SerializationAtDifferentLevels} has been implemented using a command stack containing a stack of serialized networks.
% An undo/redo triggers a deserialization of the processor network from the current location of the stack.
% While this is conceptually simple, it requires the processor network to support partial deserialization, i.e., it is not possible to remove all parts of the network and create them when deserializing since it would be too slow. 
% Instead, the deserialization is applied to the existing network, removing, adding and updating parts as necessary.
% Parts that does not change are unaffected by the deserialization.
% Details about the partial deserialization can be found in the source code.

% The partial deserialization described above is also used to support copy/paste of parts of a processor network.

\subsection{Two-Level Documentation}

Inviwo implements the documentation concept combining low level API-documentation and high level documentation in the network editor using the Doxygen system. 
The Doxygen docpage command is used in the header file of the processor, thus co-locating the documentation for developers and visualization editor users.
Each docpage provides descriptions of how to use the processor, its inports, outports and parameters. 
In practice, we use Markdown language for this purpose, since it is supported by Doxygen and thereby provides a rich way for describing the module.
The documentation itself is generated at compile time using a separate compile target.
The processor identifier, which is a unique identifier available in all processors, is used to provide a connection between the documentation and the processor.
This makes it possible to also automatically generate images of the visual representation of each processor.
These images provide a contextual overview of the processors in the documentation.
An example of API and end user documentation based on the same comments can be seen in Figure~\ref{fig:backgroundDocumentation}.
In order to reduce copy-paste errors, and to make it quick and easy to create as well as document a processor, boiler plate processor code can be generated through a python script.

%% -----------------------------------------------------------------------------------------------------------

\subsection{Cross-Layer Testing}
The Inviwo system tries to make it as easy as possible for developers to create tests across the layers of abstraction, i.e,  code and entire processor networks.
For example, the tests made for Inviwo can run on the developer's machine, they do not require a separate testing machine. 
The system currently provides two means for testing, unit tests, which focus on testing each element of the code, and regression tests, which focus on testing integration between processors, i.e., the top abstraction layer described in \sref{sec:testing}.

Module developers can add tests by creating a folder called "tests" in their module directory and putting unit tests in sub-folder "unittests" and regression tests in sub-folder "regression".
All tests organized according to this structure will automatically be added by the framework. 
Further details about the two types of testing is provided below.
%% -----------------------------------------------------------------------------------------------------------

\subsubsection{Unit Testing}
%Unit testing plays a major role in ensuring correct results by testing and guaranteeing assumptions on the expected output of some source code.
%This includes testing algorithms and methods for plausible results, special values, extrema, etc.
%It is therefore an important part of the development within Inviwo.
%The unit testing itself is based on Google Test, the C++ test framework provided by Google~\cite{whittaker2012google}.
Inviwo relies heavily on Google Test~\cite{whittaker2012google} for its unit testing and, instead of providing its own unit testing framework, tries to provide a smooth integration.
The unit tests are performed on a per-module basis.
As mentioned above, all files having cpp file-ending residing inside the unit testing folder of the module are automatically considered.
Each test source file can include one or multiple tests.
By default, unit testing of all modules is included in the build process of the system.
This means that the tests are run after a successful build and output information about the tests is shown in the integrated developer environment (IDE).
Unit tests can also be run individually if desired.

\subsubsection{Regression Testing}
While writing unit tests for each individual unit of code ensure correctness of the unit, it does not necessarily test its integration with other units.
Regression tests in Inviwo execute a whole network and compare its output with a result deemed correct.
Creating a regression test is therefore only a matter of saving the output in the processor network, commonly the canvas images, along with the workspace containing the network.
Future changes producing a different output will fail the regression test.
Thus, this is a quick and easy way of detecting if code changes have an undesirable effect on the output of existing algorithms.
While this approach does not tell exactly which unit of code caused the failure, it does tell which code changes that did.
From experience, we have found that GPU hardware might produce small numerical differences depending on driver version or manufacturer.
To resolve this, the developer can configure the regression test error threshold per output.
Note that allowing errors should be used with caution and only after thorough investigation.

\begin{figure}
  \centering
 \includegraphics[width=\linewidth, trim={0 0 0 0}]{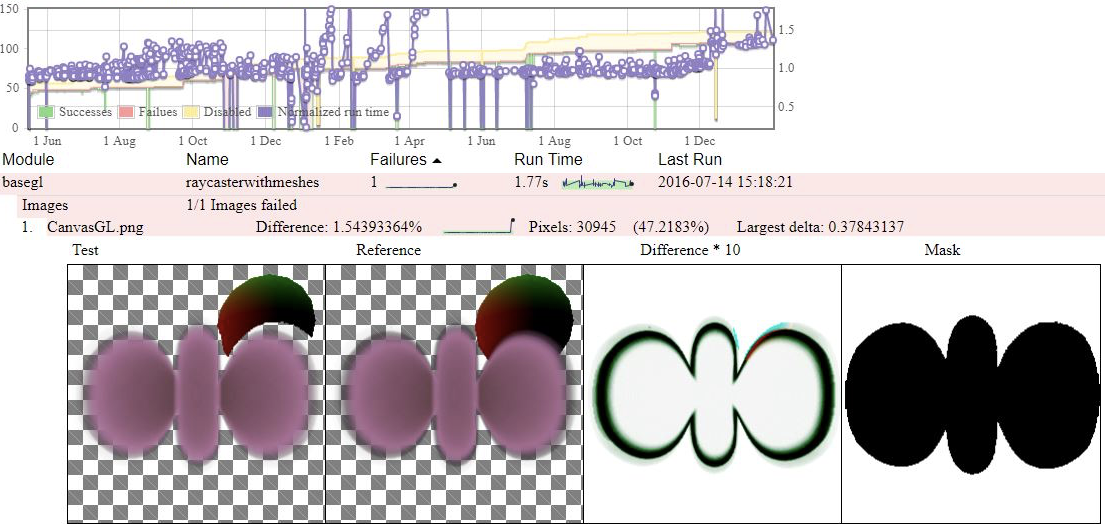}
  \caption{ \label{fig:regression} Depiction of a regression test report.
  The plot in the top provides an overview of the regression test history. 
  Test details show the new output, the reference, magnified difference between them and a binary mask of where error is above the error threshold. 
  Additional details include the runtime log and explanation of the error causing the failure.}
\end{figure}

Python scripts can be used to simulate interaction or other runtime changes. 
All Python scripts in the folder of the regression test will be executed after loading the network, meaning that all functionality exposed through the Inviwo Python integration can be utilized in a regression test.
The Inviwo regression testing environment itself is setup using Python scripts. 
The Python scripts manage the execution of the tests and also generate a report in HTML.
An example of a regression test report is depicted in \fref{fig:regression}.
Besides details about the outcome of the tests, the report allows developers to see the difference between the result image and the reference image as well as analyzing performance measurements over time.

Regression tests need to be executed on demand since they are too time-consuming to run as a build-step. 
Correctness and stability of the public repository code is ensured by utilizing a continuous integration server, which runs all tests before merging code changes. 

\subsection{Algorithm Interoperability}
Inviwo has implemented the data interface described in \sref{sec:computePlatformCommunication} using template functions for retrieving representations.
The returned representations are therefore strongly typed, which aids in the development process.
Currently, CPU, OpenGL and OpenCL computing platforms are supported.
We foresee that future heterogeneous computing platforms may integrate CPUs and GPUs into the same hardware. 
The data location will in this case be be shared among the different computing platforms.
Inviwo system takes this scenario into account by allowing the underlying data to the shared among different computing platforms.
Currently, Inviwo supports shared OpenGL and OpenCL buffers and textures.
No data transfer is in this case necessary, since they reside at the same video memory location.
There are a range of implementation aspects that need to be considered when using shared OpenCL representations. 
First, it requires that data has first been allocated in OpenGL.
Thus, if data is located on RAM it must first be transfered to OpenGL before a shared OpenCL representation can be created.
The algorithmically created converter packages described in \sref{sec:computePlatformCommunication} substantially reduces the implementation effort in this case. 
Second, OpenCL supports a limited subset of data formats compared to OpenGL.
Lower versions of OpenCL do for example not support depth textures.
Therefore, Inviwo currently throws errors when these unsupported formats are used.
Ideally, they should be converted without errors even though performance might suffer, which can be seen as a future improvement for the system.
The implemented computing platform converters makes Inviwo computing platform interoperable and allows current and future computing platforms to be used without changing existing implementations as required in \sref{sec:systemrequirements}.
\section{Application Examples}
\label{sec:applicationexamples}

Inviwo has successfully been used in numerous scientific publications within different application areas, commercial products, and university-level courses.
The scientific contributions include work on advanced volumetric illumination~\cite{SR15,Jon16,JY17}, medical and molecular visualization~\cite{kreiser2018visually,FYTL19,Kotravel2019}, transfer function design for volume rendering~\cite{JFY16,falk2017colortf}, crowd-sourcing-based user studies~\cite{ER16,EKR16,ER18}, topological analysis~\cite{Bock18, koepp19a, Jankowai2019}, as well as multi-variate data and flow visualization~\cite{Jankowai2018a, ERH16, ELHY18}.
In addition, Inviwo is to the best of our knowledge currently used by four different universities and two commercial companies.

%\todo{I would mention the Master's theses in \sref{sec:example:education} / Martin}

%Additional things to mention?
%\begin{itemize}
%  \item Compare amount of code using our framework against not using our framework. 
%    (OpenGL example, compare with Figure~2 in Voreen paper?)
%  \item Python bindings etc.
%  \item matlab, R?
%  \item Dispatching data formats
%\end{itemize}

We have selected four examples demonstrating usage of Inviwo at different usage abstraction levels as well as its computing platform interoperable capabilities.
Admittedly, some of these concepts are hard to demonstrate since they are integrated in the development process.
We start by demonstrating the core concepts in a fictitious example starting from importing data over adding additional functionality in shaders and processors to creating regression tests.
Afterward, we continue with real-world examples including large scale pathology imaging, public brain visualization exhibition, and education.

\subsection{Usage Abstraction Level Work-Flow}
This example demonstrates how the presented concepts are used in the process of creating a visualization. 
It starts on a high usage abstraction level, goes into medium and low usage abstraction levels before finishing with creating a visualization pipeline test. 
The reader is strongly encouraged to have a look at the video in the supplementary material which complements this textual description.

In order to create a volume rendering pipeline, the user starts by dragging and dropping a volume source processor set into the network editor and referring to a CT scan of a salmon.
Next, three more processors are added to the pipeline; one for providing the bounding box geometry for the volume, one for generating entry and exit points for each pixel, and a volume raycaster.
This processor renders the volume by considering the output of the other processors including the volume source.
Finally, a canvas processor is added to show the resulting image.
The volume data is then further explored by adjusting the transfer function.
To quickly get an understanding of the data flow in the pipeline, the user hovers over the outports of each processor and thereby sees their content by means of the port inspector concept.

In the next step, the user wants to clip the volume programmatically in the shader, which we refer to as medium usage abstraction level.
The raycasting shader is extended accordingly and the changes are immediately propagated to the network after saving.
The necessary parameters, i.e.\ a boolean flag and the spatial position of the clipping, are exposed via shader uniforms.
On the low usage abstraction level, the user can now add matching properties to the C++ source of the volume raycaster processor.
These properties are then used to set the uniforms in the shader.
A recompilation of the code causes the network editor to reload the network, which demonstrates the interactive development possibilities even at the low usage abstraction levels.
Moving back to high level editing in the network editor, the newly added properties are accessible in the property list and can be used to interactively adjust the clipped region.

High and medium usage abstraction levels are exposed by the Python scripting integration.
Here, the user decides to export each xy slice of the volume dataset individually which requires a script looping over all slices in the volume and writing the respective contents of the canvas to disk.

To ensure that the results remain consistent despite future changes, a regression test is created by selecting create regression test in the menu. 
This demonstrates the ability to quickly create tests at high usage abstraction levels.
The regression test is run locally and the resulting regression report is inspected.

\subsection{Applied Research Usage Example}
\label{sec:example:digitalpathology}
%This application scenario is taken from the medical domain and serves as an example for large-scale data handling.

% In histology, a field within pathology, tissues and cells are studied on a microscopic scale. 
% Tissue samples are fixated and cut into thin slices before being put on microscope glass slides.
In recent years, hospitals have started to digitize tissue samples on glass slides using scanners to obtain digital high-resolution color images of the samples~\cite{thorstenson2014wsisweden}.
Multiple digitized slide images of the same sample can be co-registered and combined into a single volume, thereby closely resembling the original block of tissue.
The challenges associated with this kind of data include dealing with large amounts of image data, ways of interacting with this data, and the ability to look at the inside of the volume, i.e., the image stack.

The visualization of the image stack is based on brick-based volume raycasting \cite{beyer2015star}, thus enabling large scale data handling of volumetric data within Inviwo.
Inviwo provides a thread pool, which is used for on demand asynchronous loading of bricked image data.
Image tiles are first loaded using the OpenSlide image library \cite{goode2013openslide} and then uploaded into a 3D texture atlas.
% The texture atlas contains the currently visible bricks as well as their direct neighbors to ensure smooth panning.
A volume, indicating the brick indices, is forwarded to the raycasting along with the texture atlas.
The existing raycasting processor was extended in order to account for RGB image data instead of scalar values.
A 2D transfer function based on color similarity \cite{falk2017colortf} enables the user to adjust opacity as well as replacement colors, which could, e.g., be used for highlighting features.

\begin{figure}[!tb]
  \centering
  \includegraphics[width=\linewidth]{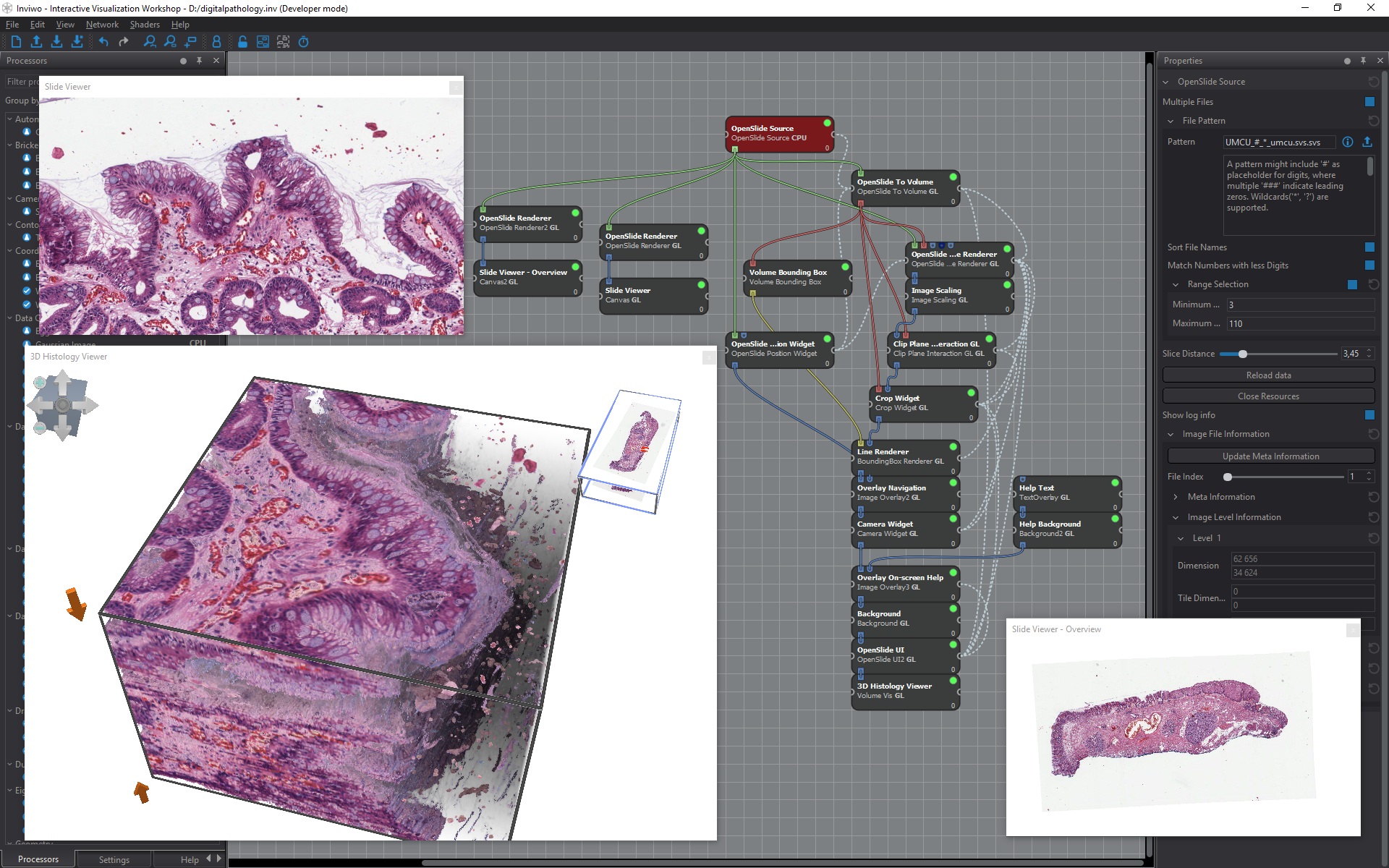}
  \caption{%
    Digital pathology:
    Visualizing a stack of 100 colored microscopic slides, each with a resolution of 63,000$\times$35,000 pixels and totaling in 110\,GiB of compressed TIFF image data.
    %The individual images have been digitized at 40x magnification, i.e.\ 0.25\,\textmu{}m per pixel, and have a resolution of 63,000$\times$35,000 pixels.
  }
  \label{fig:digitalpathology}
\end{figure}

\fref{fig:digitalpathology} shows an interactive rendering of 100 slide images as well as domain specific navigation. 
The TIFF jpeg compressed image data consumes about 100\,GiB on disk and features details like individual nuclei.
In order to address the challenges related to interacting with the stack of slide images, Inviwo provides a user interface, a minimap, and navigation widgets.
The result is a tailored application that is used by a number of pathologists located at several hospitals and was evaluted in a user study with practising pathologists \cite{FYTL19}.
This scenario has demonstrated the ability to develop at both low and high abstraction layers in Inviwo as well as its capabilities in tailoring an application to domain scientists.
%The primary use cases are examining existing datasets and evaluating the potential use of 3D histology in clinical practice.

\subsection{Public Dissemination Usage Example}
\label{sec:example:brainvis}

Taking research into public display, such as science centers and museums, poses different demands on a system compared to prototyping ideas for research.
Stability, in case the application will be running all day, seven days a week, is a an even more important requirement.
In addition, the application must be responsive and the user interface should be tailored to suit visitors without prior experience regarding the underlying algorithms~\cite{JFY16}. 

\begin{figure}
 \centering
 \includegraphics[width=\linewidth, trim={0 100px 0 0}, clip]{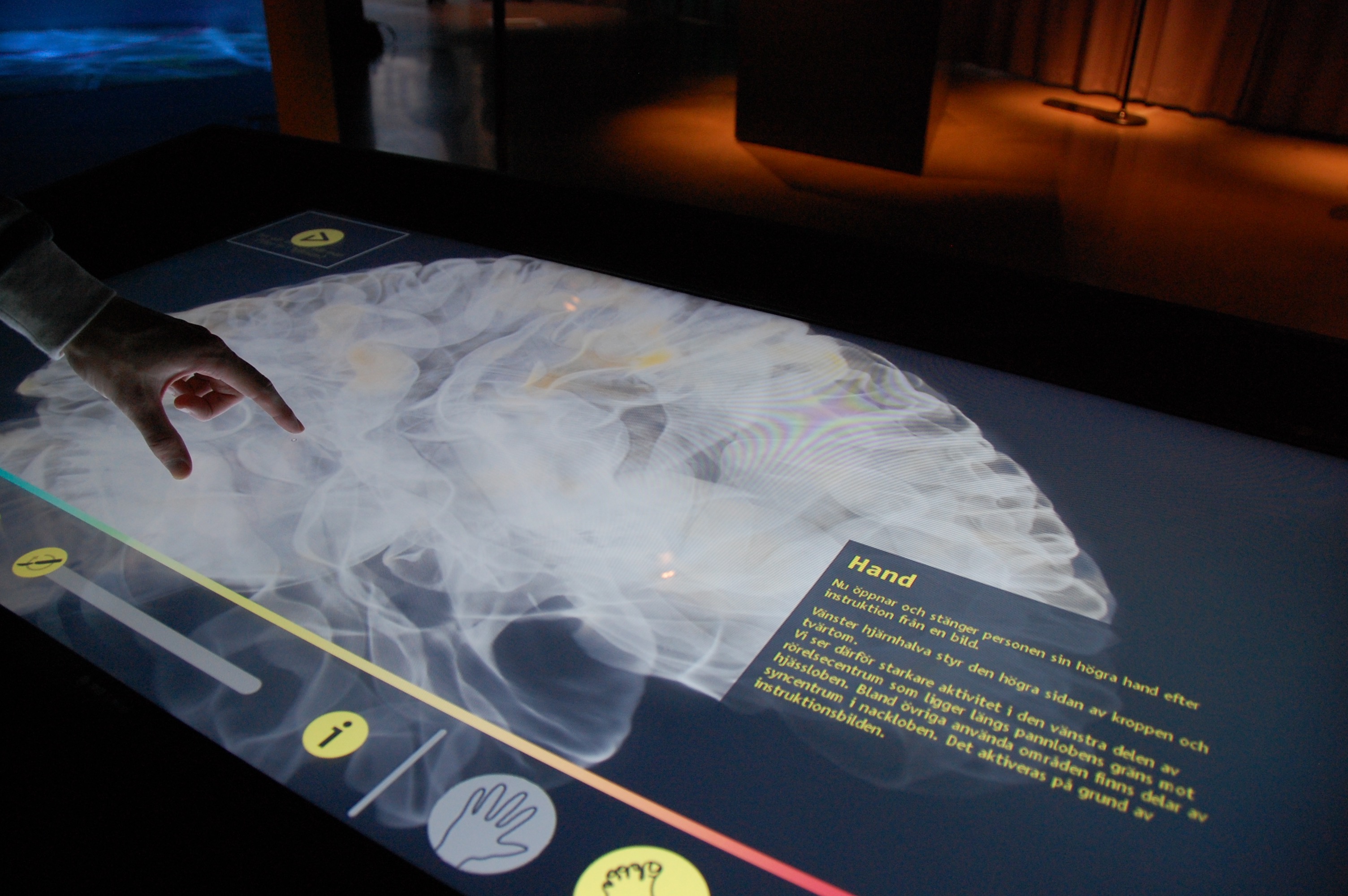}
 \caption{ \label{fig:brainvis} Public exhibition of human brain activity depicting where active areas are located in the brain when performing different tasks.}
\end{figure}

\fref{fig:brainvis} depicts a brain visualization application where the entire user interface has been tailored for public use.
The application visualizes where active areas in the brain are located while performing different tasks, such as wiggling toes or listening to music.
It uses volume visualization techniques to fuse data from magnetic resonance imaging (MRI) and functional MRI, inspired by Nguyen et al.~\cite{NEOHLFAKY10}.
The graphical user interface communicates with the processor network driving the brain visualization.
It uses the built-in processor network evaluation, touch interaction, and data management, while extending Inviwo with new modules for multi-modal volume visualization. 

In this scenario, the work-flow for creating the application was as follows.
In a first step, the visualization pipeline was set up in the network editor using readily available processors for loading MRI and fMRI neuroimaging data, data processing, and interaction.
A new volume visualization processor was created for fusing the MRI and fMRI signals.
Finally, Inviwo was used as an API in an application with a custom user interface, where the GUI elements were connected to the corresponding properties in the processor network using their identifiers.

%has been utilized to create an application where novice users of all ages can explore real-captured fMRI and MRI data. 
%The brain activity is partially based on data from the Human Connectome Project.
%One of the goals for this project was to utilize interactive visualization for educating the public about scientific data. 
%This application was created in 2015, for several exhibitions, which span a life-time of up to 10 years.

The application has been running every day during opening hours since 2015 at the Tekniska museet (Technical Museum) in Stockholm, Sweden, and the Visualiseringscenter C (Visualization Center C) in Norrk\"{o}ping, Sweden.
A big advantage of this setup has been that the actual data processing, rendering, and interaction could be fully created and adjusted on a higher level of abstraction within the Inviwo network editor.
This application scenario demonstrates the use of Inviwo  at a low level, as an API, at a high level through the visual network editor, as well as its stability, responsiveness, and user interface flexibility.

\subsection{Educational Usage Example}
\label{sec:example:education}
Inviwo has been used as the underlying platform for a number of master thesis' and is currently being used in, to our knowledge, four courses on visualization. 
In these courses, the students are primarily using the system to get an in-depth understanding of visualization algorithms by implementing them, but also to explore how common visualization algorithms work in practice. 
VTK~\cite{schroeder2006visualization}, Paraview~\cite{Ayachit2015ParaView} and VisTrails~\cite{bavoil2005vistrails} have previously been used for two of these courses.
From this experience, it was found that use of pure VTK imposed a steep learning curve and that the tree-view of data-flow in Paraview was hard for the students to understand with respect to how data is flowing in the visualization pipeline. 
VisTrails has succesfully been used for education~\cite{CGF:CGF1830} and provides a similar setup as Inviwo with respect to the visualization pipeline. 
However, VisTrails tends to expose many inports and outports per node/processor, which was confusing to the students. 
In Inviwo, the number of ports of processors are reduced by applying the processor creation guidelines presented in the appendix, which make it easier for new users of the system.
The visual debugging concepts further help the students to understand the data flows and find errors in their implementations.

As a contrast to the courses targeting visualization, Inviwo is also used in a course for physics students on Bachelor level.
The physics students design visualizations of electron structure simulation data for common tasks involving analysis of charge densities, molecular dynamics and crystal structures.
They use both the Python scripting in Inviwo as well as the high level network editor to accomplish their goals.

To summarize, Inviwo has been used by hundreds of students to both get in-depth understanding of visualization algorithms and how to use them to understand data.
The students use the system at all levels of abstraction.

%\begin{itemize}
%  \item how has Inviwo been used for eduction?
%  \begin{itemize}
%    \item courses / labs
%    \item Visualization course
%    \item InfoVis course?
%    \item add brief description on benefits using Inviwo compared to what was there before
%  \end{itemize}
%  \item cite Master's theses (\todo{cite SIGRAD only? / Martin})
%  \begin{itemize}
%    \item Torstens A-buffer: thesis, \cite{Gustafsson2017}, sigrad: \cite{GEEH17} 
%    \item Jochens work thesis: \cite{Jankowai2016}, sigrad:\cite{jochen16})
%  \end{itemize}
%\end{itemize}

%% -----------------------------------------------------------------------------------------------------------
\section{Discussion And Conclusions}
\label{sec:conclusions}
%The requirements on visualization systems today are high and require much effort to realize.
In this paper we have described how to design a computing platform interoperable visualization system with usage abstraction levels through a layered architecture.
The proposed computing platform interoperable solution allows algorithms developed for different computing platforms to be used together while still allowing developers to access the underlying computing platforms.

We presented several concepts for interactive development, debugging, documentation and testing across layers of abstraction.
More specifically, the concepts allow tasks commonly performed by a developer at lower levels, i.e., debugging in IDEs, to be performed on a visualization pipeline level.
While some of these concepts are found in other systems, we have formalized them and brought them together in one system.
It is the combination of the presented concepts that enable tailored visualization applications to be rapidly developed.
For example, since the system is computing platform interoperable, it is possible to design a port inspector using OpenGL and apply it to data residing in OpenCL.
This also demonstrates the crucial success factor in view of the amount of work that goes into creating and maintaining advanced visualization systems.

The concepts were demonstrated in the Inviwo system along with accounts for implementation choices and technical details, such as use of shared computing platforms, that need to be considered when realizing the concepts. 
Naturally, not all details can be provided in a systems paper but a comprehensive account for technical aspects is available on the Inviwo website (\url{www.inviwo.org}), where the source code is also distributed freely under the BSD license.

%building blocks provided by the core functionality and algorithms can be introduced using modules.
%Extending the framework is done using modules, which is essential for enabling the applications demonstrated in Section~\ref{sec:applicationexamples}.  

% Naturally, not all aspects of a complex visualization system can be discussed, both due to the nature of a research paper but also the imposed page limit.
% We therefore selected a set of architecture-level features that might be of interest to both developers of Inviwo as well as other frameworks.
% These features span a wide range, from low-level data structures to documentation and visual introspection, but show a sample of benefits that come with using the Inviwo framework.
% We further explained how both unit and regression tests can be added for maintaining correct and consistent results. 
% It was shown how interactive reports with visual comparison of image outputs makes it easier to understand an error and its source.

Several application examples, implemented using the Inviwo system, demonstrated the wide usage levels made available through the described concepts.
It was shown that Inviwo meets the requirements and can be used throughout the whole visualization application design process, from writing functional units, through visualization pipeline editing, to tailored application creation.

While the presented computing platform interoperable solution allows for high performance and novel utilization of each computing platform, it can also be seen as one of the limitations since algorithms will be designed for a specific computing platform and will have to be ported to devices not supporting them.
For example, algorithms using OpenCL cannot run on mobile phones.
The presented concepts have been designed and demonstrated for single-computer usage. An interesting direction for future work would be to take the demonstrated out-of-core concepts and extend them to distributed computing.

% Each example described challenges within its area and its use of Inviwo to support the development process for addressing those challenges.
% These examples show that Inviwo can be, and is, used for research, public dissemination as well as education. 
%The paper detailed how to obtain, build and extend Inviwo in order to integrate new visualization algorithms or to use the existing processors to deploy custom visualization applications. 
%To ensure a long halftime of the aspects discussed in this paper, we have described the developed technologies on a conceptual level first, before providing source code detailing how selected aspects are implemented and used. 
 
%Therefore, we have only focused on those aspects, which are relevant for visualization researchers and practitioners, in order to get started and generate their first visualizations with in Inviwo. 
% For any further information we would like to refer to the Inviwo website (\url{www.inviwo.org}), where additional resources, documentation and tutorials can be found. 
%Through the forum available on the website Inviwo users can also get into touch with the main developers and the authors of this paper.

%% -----------------------------------------------------------------------------------------------------------

%\begin{appendix}
%\section*{Appendix - Processor Creation Guidelines}

\section*{Acknowledgments}
%\acknowledgments{
This work was supported through grants from the Swedish e-Science Research Centre (SeRC), the Deutsche Forschungsgemeinschaft (DFG) under grant RO 3408/3-1 (Inviwo), the Excellence Center at Link\"oping and Lund in Information Technology (ELLIIT), the Knut and Alice Wallenberg Foundation (KAW) grant 2013-0076, and the Swedish research council grant 2015-05462. The authors further thank the Center for Medical Image science and Visualization (CMIV) at Linköping University and the Ulm University Center for Translational Imaging MoMAN for their support. The current version of Inviwo can be downloaded at www.inviwo.org.

%The Barrett’s histology dataset was kindly provided by Dr.~Marnix Jansen, UCLH and Barts, London, UK.

%We were paid by SeRC and SeRC and SeRC and SeRC and SeRC


\ifCLASSOPTIONcaptionsoff
  \newpage
\fi

\bibliographystyle{IEEEtran}
\bibliography{bibliography}

\vspace{-2em}
\begin{IEEEbiography}[{\includegraphics[width=1in,height=1.25in,clip,keepaspectratio]{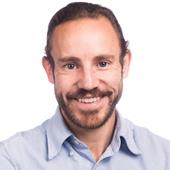}}]{Daniel J\"{o}nsson}
received the MSc and PhD degrees in
media technology in 2009 and 2016 from Link\"{o}ping
University, Sweden. Since 2017, he is a research fellow at the division for Media and
Information Technology (MIT) at the Department of Science and Technology (ITN),
Link\"{o}ping University. His current main research interests lie in the intersection between visualization and artificial intelligence.
\end{IEEEbiography}\vspace{-2em}
\begin{IEEEbiography}[{\includegraphics[width=1in,height=1.25in,clip,keepaspectratio]{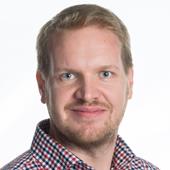}}]{Peter Steneteg} is a principle research engineer in the Scientific Visualization Group at Linköping University. He received his Ph.D.\ degree in theoretical physics from Linköping University in 2012. He is currently the Project Manager and Lead Core Developer of Inviwo.
\end{IEEEbiography}\vspace{-2em}
\begin{IEEEbiography}[{\includegraphics[width=1in,height=1.25in,clip,keepaspectratio]{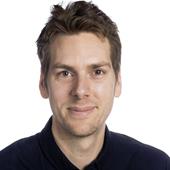}}]{Erik Sund\'{e}n} received the MSc in media technology in 2010 from Link\"{o}ping University, and worked as a research engineer between 2010 and 2015 within Scientific Visualization. Since 2015 he is Technology Manager at Norrk\"{o}ping Visualization Center C, partly run by the the division for Media and
Information Technology (MIT) at the Department of Science and Technology (ITN). He conducts productions and research projects with aims of both state-of-the-art research in visualization as well as public outreach.
\end{IEEEbiography}\vspace{-1em}
\begin{IEEEbiography}[{\includegraphics[width=1in,height=1.25in,clip,keepaspectratio]{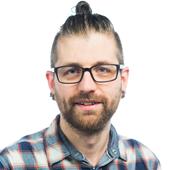}}]{Rickard Englund} is a PhD Student in Scientific Visualization at Linköping University. He received his master’s degree in media technology from Linköping University in 2014. His research interests lie within scientific visualization with a focus on interactive exploration of temporal vector fields. 
\end{IEEEbiography}\vspace{-1em}
\begin{IEEEbiography}[{\includegraphics[width=1in,height=1.25in,clip,keepaspectratio]{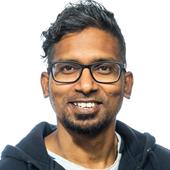}}]{Sathish Kottravel} is a PhD Student in Scientific Visualization at Linköping University. He received his master’s degree in advanced computer graphics from Linköping University in 2014. His research focus on parallel algorithms for scientific visualization. 
\end{IEEEbiography}\vspace{-1em}
\begin{IEEEbiography}[{\includegraphics[width=1in,height=1.25in,clip,keepaspectratio]{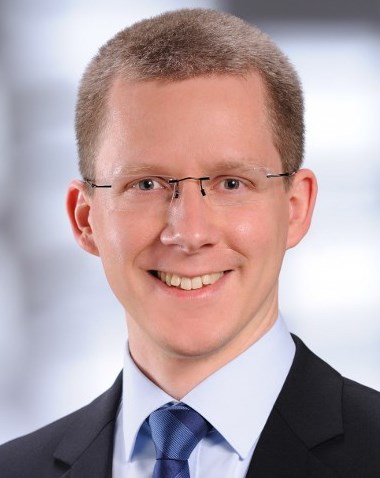}}]{Martin Falk}
is a research fellow in the Scientific Visualization Group at Linköping University. He received his Ph.D.\ degree (Dr.rer.nat.) from the University of Stuttgart in 2013. His research interests include large-scale volume rendering, visualizations in the context of pathology and systems biology, large spatio-temporal data, topological analysis, glyph-based rendering, and GPU-based simulations.
\end{IEEEbiography}\vspace{-1em}
\begin{IEEEbiography}[{\includegraphics[width=1in,height=1.25in,clip,keepaspectratio]{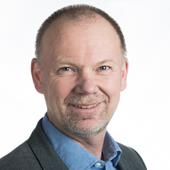}}]{Anders Ynnerman}
received a Ph.D. 1992 in physics from Gothenburg University. During the early 90s he was doing research at Oxford University, UK, and Vanderbilt University, USA. In 1996 he started the Swedish National Graduate School in Scientific Computing, which he directed until 1999. From 1997 to 2002 he directed the Swedish National Supercomputer Centre and from 2002 to 2006 he directed the Swedish National Infrastructure for Computing (SNIC). Since 1999 he is holding a chair in scientific visualization at Linköping University and he is the director of the Norrköping Visualization Center - C, which currently constitutes one of the main focal points for research and education in computer graphics and visualization in the Nordic region. The center also hosts a public arena with large scale visualization facilities. He is also one of the co-founders of the Center for Medical Image Science and Visualization (CMIV).
\end{IEEEbiography}\vspace{-1em}
\begin{IEEEbiography}[{\includegraphics[width=1in,height=1.25in,clip,keepaspectratio]{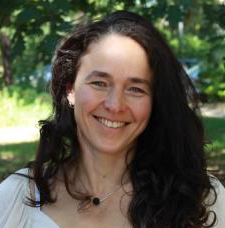}}]{Ingrid Hotz}
is a Professor in Scientific Visualization at the Linköping University in Sweden. She received her PhD degree from the Computer Science Department at the University of Kaiserslautern, Germany. Her research interests lie in data analysis and scientific visualization, ranging from basic research questions to effective solutions to visualization problems in applications. This includes developing and applying concepts originating from different areas of computer sciences and mathematics, such as computer graphics, computer vision, dynamical systems, computational geometry, and combinatorial topology.
\end{IEEEbiography}\vspace{-1em}
\begin{IEEEbiography}[{\includegraphics[width=1in,height=1.25in,clip,keepaspectratio]{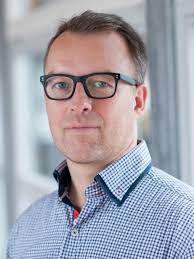}}]{Timo Ropinski} is a Professor in Visual Computing at Ulm University, Germany, where he is heading the Visual Computing Research Group. Before his time in Ulm, he was Professor in Interactive Visualization at Linköping University, Sweden. Timo holds a PhD from the University of Münster, Germany, where he also finished his Habilitation. His research interests lie in data visualization and visual data analysis. Together with his research group, Timo works on biomedical visualization techniques, rendering algorithms and deep learning models for spatial data.
\end{IEEEbiography}\vspace{-1em}

\end{document}

\appendix[Processor Creation Guidelines]
\label{sec:ProcessorGuidelines}
The concept of encapsulating an functional unit into a processor is quite straightforward, but it can sometimes be difficult to  differentiate between ports and properties.
All algorithm parameters could potentially be exposed as inports. 
However, including many ports on each processor makes the system difficult to use and understand, especially if they are of the same type since they will only differ by the name the developer gave them.

The Inviwo system uses different colors for each data types and also separate visual representations for optional and mandatory ports to alleviate this problem, see Figure~4. 
Still, there must be a balance between what is exposed through ports and what is exposed as properties.
Based on years of development and usage, we have come up with a set of guidelines for this purpose.
In general, we have experienced that simplicity outweighs flexibility since it lowers the difficulty level for using the system, reduces cluttering and thus provides better overview, and also makes it faster to use overall since fewer connections are necessary. 
More flexible solutions are mostly only required by more experienced users, who already have deeper knowledge on how to use the system and therefore can create specialized solutions for their specific needs. 
The following recommended principles thus favor simplicity and ease of use in the network over flexibility:
%We try to apply when representing an algorithm using a processor.
%The guidelines aim to make it easier to use processors in the network.

\textbf{- Use inports for mandatory parameters.}
Processing will not take place until inports are connected. Thus, a good design uses ports for mandatory parameters.
Properties require default values and generally do not put restrictions on evaluation of the processor so they should not be used for mandatory parameters.

An exception to this rule is for example file path parameters. While they are mandatory for loading files they are not suitable to use as ports since they need to be specified by the user. 
Properties can easily be exposed to the user and are therefore appropriate for file path parameters.

\textbf{- Minimize the number of inports (less than five).}
Processors with many inports are hard to reuse for other purposes since they rely on many different types of inputs.
We therefore classify processors with more than five inports as a sign of bad design of the underlying algorithm and encourage them to be broken down to smaller elements.
This also affect the higher usage abstraction levels of the system. 
Each port requires a connection to another port and it quickly becomes difficult to immediately know which port to connect to.
Furthermore, each connection adds clutter to the network and therefore reduces the users' ability to overview the pipeline.

\textbf{- Use properties when there is need for tuning by the user.}
The Inviwo network editor exposes properties through a rich set of widgets, which makes it easy for users to tune parameters. Different semantics can be used for the same type of property, i.e., four floating point values can be edited through sliders, a color widget, or spin boxes. 
Port data, on the other hand, must be come from memory or files, which is not suitable for tuning by a user.

\textbf{- Use properties for parameters with low memory transfer overhead.}
Linking (synchronizing) two properties involve a copy operation each time one of them changes.
Thus, the memory transfer overhead could become noticeable when multiple processors require a parameter to be the same and it changes often.
This memory transfer overhead is in many cases not a problem since the computation time in the pipeline widely exceeds parameter copy operation time. 
We have therefore favored properties instead of ports to reduce the number of inports in the case of for example cameras and transfer functions.
A transfer function/color map can use many control points for mapping values to colors, and might cache a lookup table, but it still does not outweigh the added simplicity of including it as a property.

If performance is crucial we recommend to use an (optional) inport instead of a property for the parameter.
In practice, this means that one first creates a new processor having the parameter as both a property and outport.
Then, instead of linking the properties between multiple processors, connect the parameter through ports. 
This solution avoids copy operations and enables the parameter to be tweaked using the property GUI.
Note that this approach should still be avoided since, as pointed out before, it makes the processor more difficult to use.
%\end{appendix}